\documentclass[lettersize,journal]{IEEEtran}
\IEEEoverridecommandlockouts

\usepackage{cite}
\usepackage{amsmath,amssymb,amsfonts}
\usepackage{algorithm}
\usepackage{color}
\usepackage{balance}
\usepackage{tabularx}  
\usepackage{booktabs}  
\usepackage{array}     
\usepackage{ragged2e}  %
\usepackage{array}
\usepackage{stfloats}
\usepackage{url}
\usepackage{verbatim}
\usepackage{upgreek}
\usepackage{graphicx,color,overpic,psfrag}
\usepackage{textcomp}
\usepackage[T1]{fontenc}
\usepackage{mathrsfs}
\usepackage{float}
\usepackage[table]{xcolor}
\usepackage[bookmarks=true, colorlinks, linkcolor=black, anchorcolor=black, citecolor=black]{hyperref}
\usepackage{latexsym}
\usepackage{bm}
\usepackage{cases}
\usepackage{fancyhdr}
\usepackage{setspace}
\usepackage{algpseudocode}
\usepackage{blkarray}
\usepackage{booktabs}
\usepackage{multirow}
\usepackage{dsfont}
\usepackage{tabularx}
\usepackage{amsfonts}
\usepackage{tikz}
\usepackage{longtable}



\usepackage{lipsum}

\newtheorem{prop}{Proposition}

	\DeclareMathAlphabet\mathbfcal{OMS}{cmsy}{b}{n}
	\ifCLASSOPTIONcompsoc
	\usepackage[caption=false,font=normalsize,labelfont=sf,textfont=sf]{subfig}
	\else
	\usepackage[caption=false,font=footnotesize]{subfig}
	\fi
	
	\allowdisplaybreaks[4]

	\def\BibTeX{{\rm B\kern-.05em{\sc i\kern-.025em b}\kern-.08em
			T\kern-.1667em\lower.7ex\hbox{E}\kern-.125emX}}
	
\begin{document}
		
		\title{Statistical CSI Acquisition For Multi-frequency Massive MIMO Systems}
		
			
			\author{Jinke Tang,~\IEEEmembership{Graduate~Student~Member, IEEE,} Li You,~\IEEEmembership{Senior~Member, IEEE,} Xinrui Gong,~\IEEEmembership{Member, IEEE,}\\Chenjie Xie,~\IEEEmembership{Graduate~Student~Member, IEEE,} Xiqi~Gao,~\IEEEmembership{Fellow, IEEE,} Xiang-Gen~Xia,~\IEEEmembership{Fellow, IEEE,} and Xueyuan Shi
			\thanks{Part of this work was presented at the IEEE WCNC 2023 \cite{10118598}.}
			\thanks{Jinke Tang,  Li You,  Chenjie Xie, and Xiqi Gao are with the National Mobile Communications Research Laboratory, Southeast University, Nanjing 210096, China, and also with the Purple Mountain Laboratories, Nanjing 211100, China  (e-mail: jktang@seu.edu.cn, lyou@seu.edu.cn,  cjxie@seu.edu.cn, xqgao@seu.edu.cn).}
			\thanks{Xinrui Gong is with the Purple Mountain Laboratories, Nanjing 211100,
				China (gongxinrui@pmlabs.com.cn).}
			\thanks{Xiang-Gen Xia is with the Department of Electrical and Computer Engineering, University of Delaware, Newark, DE 19716, USA (e-mail: xxia@udel.edu).}
			\thanks{Xueyuan Shi is with State Grid Jiangsu Electric Power Co. Ltd Ganyu Branch, State Grid Corporation of China (SGCC), Lianyungang 222100, China (email: xyshi@aa.seu.edu.cn).}
		}

			\maketitle

			\begin{abstract}

		Multi-frequency massive multi-input multi-output (MIMO) communication is a promising strategy for both 5G and future 6G systems, ensuring reliable transmission while enhancing frequency resource utilization. Statistical channel state information (CSI) has been widely adopted in multi-frequency massive MIMO transmissions to reduce overhead and improve transmission performance. In this paper, we propose efficient and accurate methods for obtaining statistical CSI in multi-frequency massive MIMO systems. First, we introduce a multi-frequency massive MIMO channel model and analyze the mapping relationship between two types of statistical CSI, namely the angular power spectrum (APS) and the spatial covariance matrix, along with their correlation across different frequency bands. Next, we propose an autoregressive (AR) method to predict the spatial covariance matrix of any frequency band based on that of another frequency band. Furthermore, we emphasize that channels across different frequency bands share similar APS characteristics. Leveraging the maximum entropy (ME) criterion, we develop a low-complexity algorithm for high-resolution APS estimation. Simulation results validate the effectiveness of the AR-based covariance prediction method and demonstrate the high-resolution estimation capability of the ME-based approach. Furthermore, we demonstrate the effectiveness of multi-frequency cooperative transmission by applying the proposed methods to obtain statistical CSI from low-frequency bands and utilizing it for high-frequency channel transmission. This approach significantly enhances high-frequency transmission performance while effectively reducing system overhead.

			\end{abstract}
			
			
			\section{Introduction}
			In the design of future 6G communication systems, the ever-increasing capacity demands and the growing number of user terminals have made spectrum scarcity a critical issue that must be addressed \cite{10118598,9237116,10054381,9369324}. To tackle these challenges, multi-frequency massive multi-input multi-output (MIMO) communication systems have emerged as a prominent research focus, leveraging diverse frequency resources \cite{10054381,zhou2021multi,10190063,10081226,9369324}. Traditional methods, such as carrier aggregation (CA), enhance bandwidth by combining multiple component carriers within the same or adjacent frequency bands, which necessitates tight synchronization and unified resource management \cite{9417091}. In contrast, multi-frequency systems utilize distinct frequency bands without requiring tight synchronization, enabling devices to independently select or simultaneously operate across multiple frequency points. This inherent flexibility allows multi-frequency systems to dynamically adapt to varying network conditions, making them particularly well-suited for heterogeneous environments \cite{10081226,9369324}.
			
			
			For instance, sub-6 GHz and millimeter-wave (mmW) frequency bands can form a dual-frequency system, and the correlation between channels at these two frequency bands has been extensively studied \cite{ali2017millimeter,ali2019spatial}. Moreover, multi-frequency communications, as a form of multi-connectivity strategy, demonstrate significant potential in enhancing the reliability of wireless networks by enabling cooperative transmissions across different frequency bands. This is typically achieved by deploying and simultaneously activating co-located antenna arrays for different bands \cite{ali2017millimeter,ali2019spatial,10075521}, where the quality of channel state information (CSI) acquisition significantly influences the system performance.

			CSI can be categorized into two types: instantaneous CSI and statistical CSI. While instantaneous CSI is widely used in massive MIMO systems for real-time signal processing, statistical CSI provides valuable long-term channel characteristics, which are essential in scenarios where instantaneous CSI is difficult to obtain or highly variable. For instance, \cite{mine}, \cite{7332961}, and \cite{9042356} utilize the spatial covariance matrix and the angular power spectrum (APS) to design pilots for different user terminals. Additionally, \cite{10026502}, \cite{9439804}, and \cite{9910031} employ APS as prior information to enhance channel estimation performance. During transmission, covariance matrices and APS can also assist in precoding, thereby improving system transmission efficiency \cite{10478820, 9847609, 9815078, 9444800, 9110855, lu2019robust1, lu2020robust}. Furthermore, in the development of integrated communication and sensing technologies, statistical CSI often serves as a channel fingerprint, which, when combined with location data, enables channel charting \cite{10430216, 10530520, jin2024i2i}. Therefore, selecting appropriate methods for acquiring statistical CSI is crucial for advancing these research areas.

			With respect to statistical CSI acquisition in multi-frequency massive MIMO systems, traditional strategies rely on training and measuring at different frequency bands. However, as the carrier frequency increases and the dimensions of the antenna array expand, acquiring statistical CSI separately for each frequency band, particularly at higher carrier frequencies, results in significantly higher overhead \cite{9351786}. To address this challenge, measurement results indicate significant similarity in channel CSI across different frequency bands \cite{zhou2021multi,9580440,9112246,9014521,9422756,10233622}. Leveraging this similarity, the statistical CSI, such as APS, of one frequency band can serve as out-of-band information to assist in obtaining the statistical CSI of other frequency bands. Specifically, when the carrier frequencies of different bands fall within the same regime (e.g., sub-6 GHz or mmWave), certain statistical CSI can be approximately considered identical \cite{9422756,10233622}. Even in scenarios with large frequency separation, such as multi-frequency systems combining sub-6 GHz and mmWave bands, statistical CSI across frequency bands still exhibits notable similarities, despite variations caused by carrier frequency differences. These similarities enable the APS of one frequency band to be applied to the transmission design of other frequency bands, thereby reducing the overhead associated with online probing to obtain APS at other frequencies \cite{10292615,ali2017millimeter,ali2019spatial}. This paper investigates efficient methods for acquiring statistical CSI in multi-frequency communication systems, based on the assumption of similar statistical CSI across frequency bands. Specifically, we focus on two key types of statistical CSI: spatial covariance matrices and APS. Using the covariance matrix of a known frequency band, we propose novel methods to predict the covariance matrices of other bands and estimate APS, thereby enabling multi-frequency cooperative transmission.

			\subsection{Prior Works}

			Extensive research has been conducted on multi-frequency systems, demonstrating that the spatial covariance matrix at an unknown frequency band can be predicted from that at a known frequency band. For instance, \cite{ali2016estimating} proposed a method combining linear extrapolation and spline interpolation for spatial covariance matrix prediction. However, this approach may not have good prediction precision in complex propagation environments. Another method, introduced in \cite{ali2019spatial}, requires the estimation of all multipath parameters, leading to prohibitively high computational complexity. Furthermore, most existing methods are designed for uniform linear arrays (ULAs), which are limited to azimuth-only scenarios. This restriction renders them unsuitable for higher-dimensional configurations, such as uniform planar arrays (UPAs), where both azimuth and elevation dimensions must be considered. Extending spatial covariance prediction to such scenarios remains an open research challenge.

			Apart from spatial covariance matrices, enhancing transmission performance often requires additional statistical parameters as prior knowledge, such as APS \cite{9444800, 9110855, lu2019robust1, lu2020robust}. Studies such as \cite{9422756} and \cite{10233622} have demonstrated that APSs across different frequency bands retain certain similarities, underscoring the role of APS as a bridge between different frequency bands \cite{10292615}. Based on these findings, we propose the following conjecture: the APS can be estimated at one frequency band and subsequently utilized to enable robust transmission at another frequency band.

			
			
	       In terms of APS estimation, existing methods predominantly focus on azimuth-only scenarios and rely on classical spectral estimation techniques \cite{stoica2005spectral}. For higher-dimensional scenarios, \cite{ali2019spatial} and \cite{park2018spatial} employed compressed sensing (CS) techniques to estimate APS. While effective, these methods involve iterative processes and matrix inversion operations, resulting in significant computational overhead. Another approach, based on deep neural networks (DNNs), was proposed in \cite{10265352}; however, it requires prior knowledge of the specific manifold of APS, which limits its applicability in general scenarios. In contrast to these limitations, this paper addresses the problem in a more general and practical setting. We focus on two-dimensional scenarios with UPAs, where both azimuth and elevation dimensions are considered. Our proposed method represents a significant advancement over existing approaches, particularly in terms of generality, efficiency, and applicability to higher-dimensional scenarios.

			\subsection{Contributions}
			
			The main contributions of this paper are as follows:
			\begin{itemize}
				\item[$\bullet$] We propose a channel model for a multi-frequency massive MIMO communication system and investigate its statistical characteristics. Our focus is on two types of statistical CSI: spatial covariance matrices and APS. We derive the mapping relationship between the spatial covariance matrix and the APS, and based on this, we analyze the similarity and correlation of channels across different frequency bands.
				
				\item[$\bullet$] Leveraging the correlation between spatial covariance matrices at different frequency bands, we propose an autoregressive (AR) method that enables the prediction of the spatial covariance matrix of one frequency band from that of another. This method relies on the construction of a two-dimensional linear autoregressive model, making it computationally efficient and straightforward to implement.
				
				\item[$\bullet$] After predicting the spatial covariance matrix from one frequency band to another, we aim to estimate the APS from the covariance matrix. We adopt the maximum entropy (ME) criterion and propose a low-complexity algorithm for APS estimation. Given the similarity of the APS across different frequency bands, the proposed method achieves high-resolution APS estimation, thereby facilitating efficient multi-frequency cooperative transmission.
			\end{itemize}

			\emph{Outline:} The rest of this paper is structured as follows. In Section~\ref{Section2}, we present the multi-frequency channel model and establish the relationship among spatial covariance matrices of different frequency bands. Section~\ref{section3} introduces an AR method for predicting spatial covariance matrices across various frequency bands. In Section~\ref{section4}, we propose an ME-based method to estimate the APS with high resolution. Simulation results are provided in Section~\ref{section5}, where we evaluate the performance of the proposed methods. Finally, the paper is concluded in Section~\ref{section6}.
			
			\emph{Notations:} In this paper, bold lowercase (uppercase) letters are used to denote column vectors (matrices). Other notations are provided in Table \ref{notations}.

		\begin{table}[!t]
			\centering
			\caption{Notations}\label{notations}
			\renewcommand\arraystretch{1.5}
			\footnotesize
			\setlength{\tabcolsep}{3pt} 
			\definecolor{lightblue}{rgb}{0.93,0.95,1.0}
			
			\begin{tabularx}{\linewidth}{>{\RaggedRight}p{3cm}>{\RaggedRight}X}
				\toprule
				Parameter & Description\\
				\midrule
				\rowcolor{lightblue}
				$\bar \jmath  = \sqrt { - 1} $ & Imaginary unit \\  
				${\bf{0}}$ & All-zeros vector (matrix) \\
				\rowcolor{lightblue}
				${\bf{I}}$ & Identity matrix \\
				${( \cdot )^{\rm T}}$, ${( \cdot )^{\rm H}}$, ${( \cdot )^*}$ & Transpose, conjugate-transpose, conjugate operations \\
				\rowcolor{lightblue}
				${[ \cdot ]_{a,b}}$ & Element at the $a$th row and $b$th column of the matrix \\    
				$\langle {\bf{a}},{\bf{b}}\rangle$ & ${{\bf{a}}^{\rm H}}{\bf{b}}$ \\
				\rowcolor{lightblue}	
				${\langle {\bf{a}},{\bf{b}}\rangle _{\bf{R}}}$ & ${{\bf{a}}^{\rm H}}{\bf{Rb}}$ \\
				${\left\langle c \right\rangle _N}$ & Modulo-$N$ operation of the constant $c$ \\
				\rowcolor{lightblue}
				$\left\lfloor x \right\rfloor$ ($\left\lceil x \right\rceil$) & The largest (smallest) integer smaller (greater) than or equal to $x$ \\
				$\otimes$ & Kronecker product \\	
				\rowcolor{lightblue}
				${\rm{diag}}\{  \cdot \} $ & Diagonal matrix with the vector argument along its main diagonal \\
				${\rm {tr}}\{  \cdot \}$ & Matrix trace operation \\
				\rowcolor{lightblue}	
				${{\mathbb{R}}^{M \times N}}$ & ${M \times N}$ dimensional real space \\
				${{\mathbb{C}}^{M \times N}}$ & ${M \times N}$ dimensional complex space \\
				\rowcolor{lightblue}	
				${\mathbb{E}}\{  \cdot \} $ & Expectation operation \\
				${\cal F}\left\{ {f(a)} \right\}(b)$ & The Fourier transform (FT) of the function ${f(a)}$ at $b$ \\
				\rowcolor{lightblue}
				${\cal F}^{-1}\left\{ {f(a)} \right\}(b)$ & The inverse FT (IFT) of the function ${f(a)}$ at $b$ \\
				$|| \cdot |{|_p}$ & $p$-norm \\
				$|| \cdot |{|_F}$ & Frobenius norm \\
				\rowcolor{lightblue}
				$\delta ( \cdot )$ & Dirac delta function \\
				${\cal A} \times {\cal B}$ & The Cartesian product of the sets ${\cal A}$ and ${\cal B}$ \\
				\bottomrule
			\end{tabularx}
		\end{table}

			\section{Multi-frequency Massive MIMO System}\label{Section2}
		In this section, we begin by presenting a multi-frequency massive MIMO system equipped with UPAs. Following this, we focus on two types of statistical CSI, i.e., APS and spatial covariance matrices, to analyze the similarities of these statistics across different frequency bands.
			
			\subsection{System Configuration and Channel Model}
\begin{figure}[htbp]
				\centering
				\includegraphics[width=0.7\linewidth]{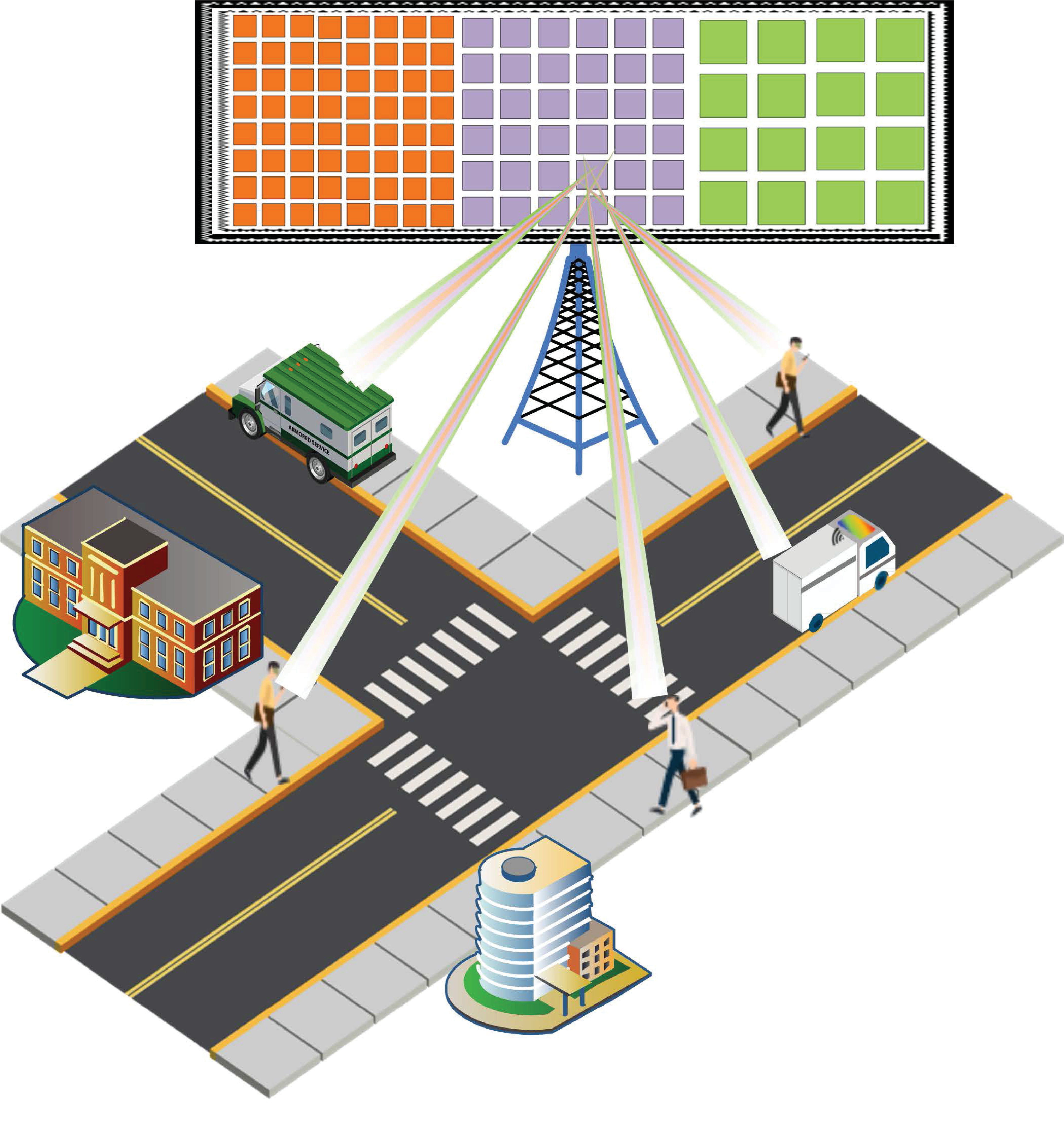}
				\caption{A multi-frequency massive MIMO system with several co-located UPAs at various frequency bands.} 
				\label{multimodel}
			\end{figure}

		We consider a multi-frequency massive MIMO system, as shown in Fig. \ref{multimodel}, where several sets of antenna arrays, operating at different frequencies, are represented by different colors in the figure. All the arrays are co-located and aligned at the same base station (BS), providing services to multiple user terminals simultaneously. In practice, the frequencies of the system, although different, may belong to the same scattering environment. Under the far-field assumption, which ensures that the propagation paths are approximately planar waves when they reach the antenna arrays, the angle of departure and angle of arrival are consistent across the arrays. This results in similar propagation parameters in the angle domain for all carrier frequencies, as demonstrated in \cite{9422756,10233622}.

			With respect to the system in Fig. \ref{multimodel}, corresponding to the frequency ${f_i}$, we assume the BS is equipped with a set of ${N_i} \times {N_i}$ UPAs, with the wavelength and inter-antenna spacings denoted as ${\lambda _i}$ and $d_i$, respectively. In this case, we use $\varphi$ and $\theta$ to represent the azimuth angle and the elevation angle at the BS side, respectively. We then define the parameters $u \buildrel \Delta \over = \sin (\theta )\cos (\varphi )$ and $v \buildrel \Delta \over = \cos (\theta )$. Consequently, the pair $(u,v)$ can be used to represent the incident direction instead of $(\theta ,\varphi )$.


			According to the ray-tracing method \cite{tse2005fundamentals,you2015pilot}, we adopt a geometric channel model, where the BS serves several single-antenna users. Therefore, at $f_i$, the channel between the $k$-th user and the BS can expressed as
			\begin{equation}\label{gk}
				{{\bf{g}}_{k,{f_i}}} = \int_{ - 1}^1 {\int_{ - 1}^{ 1} {{\alpha _{k,{f_i}}}(u,v){{\bf{v}}_{{f_i}}}(u,v)dudv} }   \in  {{\mathbb{C}}^{N_i^2 \times 1}},
			\end{equation}
			where ${{\alpha _{k,{f_i}}}(u,v)}$ and ${{{\bf{v}}_{{f_i}}}(u,v)} \in {{\mathbb{C}}^{N_i^2 \times 1}}$ represent the complex-valued channel gain and the BS response vector at the direction $(u,v)$, respectively. With the following definitions
			\begin{equation}
				{{\bf{v}}_{x,{f_i}}}(v) = \left[ {1,{e^{ - \bar \jmath 2\pi \frac{{{d_i}}}{{{\lambda _i}}}v}}, \cdots ,{e^{ - \bar \jmath 2\pi ({N_i} - 1)\frac{{{d_i}}}{{{\lambda _i}}}v}}} \right]^T,
			\end{equation}
			\begin{equation}
				{{\bf{v}}_{z,{f_i}}}(u) = \left[ {1,{e^{ - \bar \jmath 2\pi \frac{{{d_i}}}{{{\lambda _i}}}u}}, \cdots ,{e^{ - \bar \jmath 2\pi ({N_i} - 1)\frac{{{d_i}}}{{{\lambda _i}}}u}}} \right]^T,
			\end{equation}
			the steering vector corresponding to any direction $(u,v)$ at the BS can then be calculated by \cite{tse2005fundamentals}
			\begin{equation}\label{vtheta2d}
				{{\bf{v}}_{{f_i}}}(u,v) = {{\bf{v}}_{z,{f_i}}}(u) \otimes {{\bf{v}}_{x,{f_i}}}(v).
			\end{equation}

			
			We normalize the power of received paths for easier analysis, and assume that the channel coefficients at different propagation directions are uncorrelated \cite{you2015pilot}, that is, 
			\begin{equation}\label{Stheta}
				{\mathbb{E}}\{ {\alpha _{k,{f_i}}}(u,v)\alpha _{k,{f_i}}^*(u',v')\}  = {S_{k,f_i}}(u,v)\delta (u-u',v - v'),
			\end{equation}
			where ${S_{k,{f_i}}}(u,v) \buildrel \Delta \over = {\left| {{\alpha _{k,{f_i}}}(u,v)} \right|^2}$ represents the power of the direction $(u,v)$. This function describes the power distribution in the 2D angle domain, which is referred to as the APS. We use ${\cal C}$ to denote the set of all carrier frequencies. According to \cite{9422756,10233622}, the channels at different frequency bands exhibit similarities in the angle domain, i.e., for any ${f_i},{f_j} \in {\cal C}$ and ${f_i} \ne {f_j}$, we have ${S_{k,{f_i}}}(u,v) \approx {S_{k,{f_j}}}(u,v)$. Therefore, APS can be regarded as frequency-independent. We omit the subscript $f_i$ and directly use ${S_k}(u,v)$ to represent the APS of all frequency bands.
			
			\subsection{Analysis on Statistical CSI}
			
			In this subsection, we primarily analyze two types of statistical CSI, namely spatial covariance matrices and APS.
			
			Based on the channel model in (\ref{gk}), at ${f_i}$, we define 
			\begin{align}\label{Rk}
				{{\bf{R}}_{k,{f_i}}} &= {\mathbb{E}}\left\{ {{{\bf{g}}_{k,{f_i}}}{\bf{g}}_{k,{f_i}}^{\rm {H}}} \right\}\notag\\
				&= \int_{ - 1}^1 {\int_{ - 1}^{ 1} {{S_k}(u,v){{\bf{v}}_{{f_i}}}(u,v){\bf{v}}_{{f_i}}^{\rm{H}}(u,v)dudv} },
			\end{align}
			which is the spatial covariance matrix of ${\bf{g}}_{k,{f_i}}$. Specifically,  ${{\bf{R}}_{k,{f_i}}}$ is composed of the covariance coefficients between any two antenna elements in the UPA. From the wide-sense stationarity (WSS) across the antenna elements \cite{mine},  for the antenna elements at the $a$-th row, $b$-th column and the $c$-th row, $d$-th column of the UPA, the covariance coefficient between them can be expressed as
			\begin{equation}\label{decorr}
				{r_{k,{f_i}}}(b - d, a- c) \buildrel \Delta \over =  {\mathbb{E}}\left\{ {{{[{{\bf{g}}_{k,{f_i}}}]}_{(a - 1){N_i} + b}}{{[{\bf{g}}_{k,{f_i}}^{\rm {H}}]}_{(c - 1){N_i} + d}}} \right\}.
			\end{equation} 
			Thus, the elements in ${{\bf{R}}_{k,{f_i}}}$ are essentially the values of ${r_{k,{f_i}}}(m,n)$, i.e.,
			\begin{equation}\label{rmn}
				{r_{k,{f_i}}}(m,n) = \int_{ - 1}^1 {\int_{ - 1}^1 {{S_k}(u,v){e^{\bar \jmath 2\pi \frac{{{d_i}}}{{{\lambda _i}}}(mu{\rm{ + }}nv)}}dudv} },
			\end{equation}
			where $m, n \in \{ 0, \pm 1, \cdots , \pm ({N_i} - 1)\} $.

			
			In practical systems, ${d_i}/{\lambda _i} \le 0.5$ always holds at $f_i$ to avoid spatial aliasing \cite{stoica2005spectral}. Under this constraint, as the carrier frequency $f_i$ increases, the wavelength $\lambda_i$ decreases, enabling antenna panels of the same size to accommodate higher-dimensional antenna arrays. Consequently, in most multi-frequency massive MIMO systems, higher carrier frequencies are associated with higher-dimensional antenna arrays \cite{10118598}. Fig. \ref{multimodel} illustrates an example of a triple-frequency massive MIMO system, where the carrier frequencies from right to left are $f_1$, $f_2$, and $f_3$, satisfying ${f_1} < {f_2} < {f_3}$, and the UPA dimensions satisfy ${N_1} < {N_2} < {N_3}$. Therefore, at $f_1$, the antenna dimension is relatively lower, and the overhead required to obtain statistical CSI for $f_1$ is also relatively lower compared to the other bands. In this case, for any ${f_i} \in {\cal C}$ where ${f_i} > {f_1}$, predicting ${{\bf{R}}_{k,{f_i}}}$ from ${{\bf{R}}_{k,{f_1}}}$ instead of directly measuring ${{\bf{R}}_{k,{f_i}}}$ provides an efficient way to reduce system overhead.



		   In the subsequent sections, the lowest carrier frequency in the system is denoted as $f_1$. We use $f_1$ as the reference frequency to predict the spatial correlation matrix ${{\cal R}_{k,{f_i}}}$ of any $f_i^{} \in {\cal C}\backslash \{ f_1^{}\}$, given the known ${R_{k,{f_1}}}$. From (\ref{rmn}), for all ${f_i} \in {\cal C}$, ${r_{k,{f_i}}}(m,n) = {r_{k,{f_1}}}\left( {\frac{{{d_i}{\lambda _1}}}{{{d_1}{\lambda _i}}}m,\frac{{{d_i}{\lambda _1}}}{{{d_1}{\lambda _i}}}n} \right)$ holds. This relationship explicitly demonstrates the dependence of the spatial correlation on both the reference frequency $f_1$ and the target frequency $f_i$. To establish a unified representation of ${r_{k,{f_i}}}(m,n)$ for different ${f_i} \in {\cal C}$, we define a two-dimensional function $\rho \left( {x,y} \right)$ for $x,y \in {\mathbb{R}}$ as
		   \begin{equation}\label{rho}
		   \rho \left( {x,y} \right) = \int_{ - 1}^1 {\int_{ - 1}^1 {{S_k}(u,v){e^{\bar \jmath 2\pi \frac{{{d_1}}}{{{\lambda _1}}}(xu{\rm{ + }}yv)}}dudv} }.
		   \end{equation}
		   With this function, for any ${f_i} \in {\cal C}$, we have ${r_{k,{f_i}}}\left( {m,n} \right) = \rho \left( {{d_i}{\lambda _1}m/{d_1}{\lambda _i},{d_i}{\lambda _1}n/{d_1}{\lambda _i}} \right)$. Hence, all the elements in ${{\bf{R}}_{k,{f_i}}}$ for different ${f_i} \in {\cal C}$ can be represented by the values of $\rho \left( \cdot \right)$. For any ${f_i} \in {\cal C}$ and ${f_i} > {f_1}$, predicting ${{\bf{R}}_{k,{f_i}}}$ from ${{\bf{R}}_{k,{f_1}}}$ is equivalent to predicting the values of $\rho \left(\cdot \right)$ from a known range to an unknown range. On this basis, several covariance matrix prediction methods have been proposed. Existing prediction methods can be categorized into two types: one directly obtains the elements in ${{\bf{R}}_{k,{f_i}}}$ by predicting the values of the function $\rho \left( {x,y} \right)$ \cite{ali2016estimating, ali2017millimeter}, while the other estimates the APS, i.e., ${S_k}(u,v)$, from ${{\bf{R}}_{k,{f_1}}}$ first and then calculates ${{\bf{R}}_{k,{f_i}}}$ by substituting the estimated APS into (\ref{rmn}) \cite{park2018spatial, ali2019spatial, 10265352}. Compared to the latter, the former method skips the estimation of APS and is therefore easier to implement. Thus, we propose a direct covariance matrix prediction method in the next section. It is important to note that we aim to establish a mapping relationship between the spatial covariance matrices of different frequency bands within the same multi-frequency system. Unlike traditional channel prediction approaches, the proposed method relies on the intrinsic statistical characteristics of the channel and the frequency-independent property, rather than frequent real-time measurements or channel reciprocity.

			\section{AR Method for Convariance Matrix Prediction}\label{section3}

			In this section, we introduce an AR method for predicting covariance matrices at other frequency bands based on the covariance matrix of a known frequency band.
			
			From (\ref{rho}), any elements in ${{\bf{R}}_{k,{f_i}}}$ can be represented by the values of $\rho \left(  \cdot  \right)$, and the covariance matrix prediction can be converted to predicting the values of $\rho \left(  \cdot  \right)$ from a known range to an unknown range. To facilitate analysis, we illustrate the proposed method using two frequencies ${f_1}$ and ${f_2}$ in the multi-frequency system as an example. Assuming ${f_1} < {f_2}$ and ${N_1} < {N_2}$, we define $\mu  = \frac{{{d_2}{\lambda _1}}}{{{d_1}{\lambda _2}}}$ and two sets ${{\cal S}_{{\rm{ori}}}} = \{ 0, \pm 1, \cdots , \pm ({N_1} - 1)\} $, ${{\cal S}_{{\rm{est}}}} = \{ 0, \pm \mu, \cdots , \pm \mu ({N_2} - 1)\}$. Using the function ${\rho}(\cdot)$ in (\ref{rho}), elements in ${{\bf{R}}_{k,{f_1}}}$ correspond to ${\rho}(m,n)$, where the variables $m,n \in {{\cal S}_{\rm{ori}}}$, while elements in ${{\bf{R}}_{k,{f_2}}}$ correspond to ${\rho} (\mu m,\mu n)$, where the variables $\mu m,\mu n \in {{\cal S}_{\rm{est}}}$. Therefore, predicting ${{\bf{R}}_{k,{f_2}}}$ from ${{\bf{R}}_{k,{f_1}}}$ is equivalent to predicting the values of ${\rho} (\mu m,\mu n)$ for $(\mu m,\mu n) \in {{\cal S}_{{\rm{est}}}} \times {{\cal S}_{{\rm{est}}}}$ from the values of ${\rho} ( m, n)$, where $(m,n) \in {{\cal S}_{{\rm{ori}}}} \times {{\cal S}_{{\rm{ori}}}}$.

            When $\mu ({N_2} - 1) < {N_1}$, i.e., ${{\cal S}_{\rm{est}}}$ lies within the range $[1 - {N_1},{N_1} - 1]$, unknown elements in ${{\bf{R}}_{k,{f_2}}}$ can be predicted by interpolation from the elements in ${{\bf{R}}_{k,{f_1}}}$. Conversely, when $\mu ({N_2} - 1) > {N_1}$, i.e., ${{\cal S}_{\rm{est}}}$ extends beyond $[1 - {N_1},{N_1} - 1]$, the corresponding ${\rho}(\mu m, \mu n)$ for any $|\mu m|, |\mu n|> {N_1} - 1$ must be predicted by extrapolation. While the method in \cite{ali2016estimating} achieves accurate results for interpolation, extrapolation cases still offer significant room for performance improvement. Specifically, these methods fail to fully utilize all known values in ${{\bf{R}}_{k,{f_1}}}$, relying only on the edge values of the known range for extrapolation. This leads to less accurate predictions, as the edge values alone may not adequately capture the underlying trends of the data. Inspired by the AR model in \cite{burg1975maximum}, we propose an AR-based approach that overcomes this limitation by leveraging all known values in ${{\bf{R}}_{k,{f_1}}}$.

			\subsection{AR Model for Multi-frequency Massive MIMO systems}
			We begin with a special case where $\frac{d_1}{\lambda_1} = \frac{d_2}{\lambda_2} = 0.5$ and $\mu = 1$. In this scenario, ${{\cal S}_{{\rm{ori}}}} \subseteq {{\cal S}_{{\rm{est}}}}$, and all elements in ${{\bf{R}}_{k,{f_1}}}$ are included in ${{\bf{R}}_{k,{f_2}}}$. Therefore, the problem of predicting ${{\bf{R}}_{k,{f_2}}}$ reduces to estimating the values of all elements in ${{\bf{R}}_{k,{f_2}}}$ that are not already present in ${{\bf{R}}_{k,{f_1}}}$. Defining ${{\cal S}_p} = \{ (1 - {N_2}),(2 - {N_2}), \cdots , - {N_1}\} $ and ${{\cal S}_q} = \{ {N_1},{N_1} + 1, \cdots ,{N_2} - 1\} $, the elements in ${{\bf{R}}_{k,{f_2}}}$ to be predicted correspond to ${\rho}(m,n)$, where $m, n \in {{\cal S}_p} \cup {{\cal S}_q}$.

			By representing the value of ${\rho}(m,n)$ as a linear combination of other values, we have
			\begin{equation}\label{2dARd1ys}
			{\rho}(m,n) =  - \sum\limits_{q = 1}^{N_1-1} {\sum\limits_{l = 1}^{N_1-1} {{a_1}} } (q,l){\rho}(m - q,n - l) + {\sigma ^2}\delta (m,n).
			\end{equation}
			With (\ref{2dARd1ys}), it is evident that the elements ${\rho}(m,n)$, where $m,n \in {{\cal S}_q}$, can be obtained through autoregressive linear operations. Therefore, the expression in (\ref{2dARd1ys}) can be referred to as the AR model.

			Normalize the coefficient ${a_1}$ such that ${b_1}(q,l) = {a_1}(q,l)/{\sigma ^2}$, and the AR model can then be expressed in the normalized form
			\begin{equation}\label{2dARd1}
			\frac{{{\rho}(m,n)}}{{{\delta ^2}}} + \sum\limits_{q = 1}^{{N_1} - 1} {\sum\limits_{l = 1}^{{N_1} - 1} {{b_1}} } (q,l){\rho}(m - q,n - l) = \delta (m,n).
			\end{equation}
			Define the set ${{\cal S}_{{\rm{AR}}}} = \{ (m,n)|m,n = 1,2, \cdots ,{N_1} - 1\} $ and ${{\bar {\cal S}}_{{\rm{AR}}}} = {{\cal S}_{{\rm{AR}}}} \cup \{ (0,0)\} $. By listing the AR expressions for all $(m,n) \in {{\bar {\cal S}}_{{\rm{AR}}}}$, the following equation set can be obtained:
			\begin{equation}\label{Rb=1}
			{{{\bf{W}}_{\rm{\rho }}}}{{\bf{b}}_1} = \left[ {\begin{array}{*{20}{c}}
				1\\
				{\bf{0}}
				\end{array}} \right],
			\end{equation}
			where the coefficient matrix ${{{\bf{W}}_{\rm{\rho }}}} \in {{\mathbb{C}}^{|{{\bar {\cal S}}_{{\rm{AR}}}}| \times |{{\bar {\cal S}}_{{\rm{AR}}}}|}}$ is composed of the elements in ${{\bf{R}}_{k,{f_1}}}$, and the vector to be solved for is
			\begin{equation}
			\begin{array}{*{20}{l}}
			{{\bf{b}}_1} = &[{{b_1}(0,0)}\quad{{b_1}(1,1)}\quad \cdots \quad{{b_1}({N_1} - 1,1)}\quad \cdots \\&{{b_1}({N_1} - 1,{N_1} - 1)}\quad \cdots \quad{{b_1}(1,{N_1} - 1)}]^T.
			\end{array}
			\end{equation}
			Thus, solving the AR model can be transformed into solving for the coefficient vector ${{\bf{b}}_1}$. It is evident that the solution to ${{\bf{b}}_1}$ can be directly obtained using the least-squares method. However, when the matrix dimension is large, the inversion operation required for solving (\ref{Rb=1}) becomes computationally intensive. Instead, we propose the following proposition to solve for ${{\bf{b}}_1}$ efficiently.

			\begin{prop}\label{prop1}
				Define a set of vectors $\{ {{\bf{v}}_1}$, ${{\bf{v}}_2}$, $\cdots$, ${{\bf{v}}_{|{{\bar {\cal S}}_{{\rm{AR}}}}|}}\} $ in  ${{\mathbb{C}}^{|{{\bar {\cal S}}_{{\rm{AR}}}}| \times 1}}$, and any two of them satisfy the following weighted orthogonality, 
				\begin{equation}\label{vRv}
					{\langle {{\bf{v}}_i},{{\bf{v}}_j}\rangle _{{{{\bf{W}}_{\rm{\rho }}}}}} \buildrel \Delta \over = {\bf{v}}_i^{\rm {H}}{{{\bf{W}}_{\rm{\rho }}}}{{\bf{v}}_j} = \left\{ {\begin{array}{*{20}{c}}
							0&{i \ne j}\\
							1&{i = j}
						\end{array},} \right.
				\end{equation}
				where ${\langle {{\bf{v}}_i},{{\bf{v}}_j}\rangle _{{{{\bf{W}}_{\rm{\rho }}}}}}$ denotes the weighted inner product with weighting matrix ${{\bf{W}}_{\rm{\rho }}}^{}$ of the two vectors ${{\bf{v}}_i}$ and ${{\bf{v}}_j}$. Then the AR coefficient vector ${{\bf{b}}_1}$ can be calculated by
				\begin{equation}\label{b1=vv}
					{{\bf{b}}_1} = \sum\limits_{i = 1}^{|{{\bar {\cal S}}_{{\rm{AR}}}}|} {v_i^*(0,0)} {{\bf{v}}_i}.
				\end{equation}
			\end{prop}
			\emph{Proof}: See Appendix A.
			
			Therefore, the key to solving ${\bf{b}}_1$ is to identify a set of weighted orthogonal vectors with respect to the weighting matrix ${{{\bf{W}}_{\rm{\rho }}}}$. Considering that the Gram-Schmidt algorithm is a method for computing an orthogonal basis, we propose a similar approach in the following. However, instead of the vector orthogonalization operation in the original Gram-Schmidt algorithm, we replace it with a weighted orthogonalization based on ${{\bf{W}}_{\rm{\rho }}}$. Considering $\{ {{\bf{a}}_1},{{\bf{a}}_2}, \cdots ,{{\bf{a}}_{|{{\bar {\cal S}}_{{\rm{AR}}}}|}}\} $, which denotes a set of linearly independent vectors in $\bm{\Omega} $, let
			\begin{equation}\label{proc1}
			{\bf{v}}_1^{(0)} = \frac{{{{\bf{a}}_1}}}{{||{{\bf{a}}_1}||}_2}.
			\end{equation}
			Then, for $i = 2,3, \cdots ,|{{\bar {\cal S}}_{{\rm{AR}}}}|$,
			\begin{equation}
			{\bf{v}}_i^{(0)} = {{\bf{a}}_i} - \sum\limits_{l = 1}^{i - 1} {\frac{{{{\langle {{\bf{a}}_i},{\bf{v}}_l^{(0)}\rangle }_{{{{\bf{W}}_{\rm{\rho }}}}}}}}{{{{\langle {\bf{v}}_l^{(0)},{\bf{v}}_l^{(0)}\rangle }_{{{{\bf{W}}_{\rm{\rho }}}}}}}}} {\bf{v}}_l^{(0)}.
			\end{equation}
			After obtaining the above-mentioned set of vectors satisfying (\ref{vRv}), the normalized form can be achieved by
			\begin{equation}\label{vi}
			{{\bf{v}}_i} = \frac{{{\bf{v}}_i^{(0)}}}{{{{\langle {\bf{v}}_i^{(0)},{\bf{v}}_i^{(0)}\rangle }_{{{{\bf{W}}_{\rm{\rho }}}}}}}}.
			\end{equation}
			Substituting (\ref{vi}) into (\ref{b1=vv}), we can obtain the value of ${{\bf{b}}_1}$, and the AR model in (\ref{2dARd1}) can then be solved.

			
			\subsection{Covariance Matrix Prediction with the AR model}
			With the AR model in (\ref{2dARd1}), all the values of ${\rho}(m,n)$, where $(m,n) \in {{\cal S}_q} \times {{\cal S}_q}$, can be predicted autoregressively. However, to predict all the unknown values in ${{\bf{R}}_{k,{f_2}}}$, three other cases also need to be considered. For all ${\rho}(m,n)$, where $(m,n) \in {{\cal S}_p} \times {{\cal S}_q}$, the AR prediction model can be expressed as
			\begin{equation}\label{2dARd2ys}
			{\rho}(m,n) = -\sum\limits_{q = 1}^{N_1-1} {\sum\limits_{l = 1}^{N_1-1} {{a_2}}} (q,l){\rho}(m + q,n - l) + {\sigma ^2}\delta (m,n).
			\end{equation}
			For all ${\rho}(m,n)$, where $(m,n) \in {{\cal S}_p} \times {{\cal S}_p}$, the AR prediction model can be expressed as
			\begin{equation}\label{2dARd3ys}
			{\rho}(m,n) = -\sum\limits_{q = 1}^{N_1-1} {\sum\limits_{l = 1}^{N_1-1} {{a_3}}} (q,l){\rho}(m + q,n + l) + {\sigma ^2}\delta (m,n).
			\end{equation}
			For all ${\rho}(m,n)$, where $(m,n) \in {{\cal S}_q} \times {{\cal S}_p}$, the AR prediction model can be expressed as
			\begin{equation}\label{2dARd4ys}
			{\rho}(m,n) = -\sum\limits_{q = 1}^{N_1-1} {\sum\limits_{l = 1}^{N_1-1} {{a_4}}} (q,l){\rho}(m - q,n + l) + {\sigma ^2}\delta (m,n).
			\end{equation}
			All the coefficients ${{a_2}} (q,l)$, ${{a_3}} (q,l)$, and ${{a_4}} (q,l)$ can be solved using a method similar to that for solving ${{a_1}} (q,l)$. Additionally, considering the conjugate symmetry of the covariance elements, ${a_1}(q,l)$ and ${a_3}(q,l)$ are conjugate, as are ${a_2}(q,l)$ and ${a_4}(q,l)$. With the proposed AR models, unknown elements in ${{\bf{R}}_{k,{f_2}}}$ can all be estimated through linear extrapolations, thereby avoiding the need for all-band training and measurements.

			In a more general case where $\mu \ne 1$, the covariance elements ${\rho}(\mu m,\mu n)$ can be estimated using the proposed AR method, provided that $\mu m$ and $\mu n$ are integers. The remaining ${\rho}(\cdot)$ values can subsequently be interpolated using the interpolation method in \cite{ali2016estimating} based on these estimated elements. This approach enables the prediction of covariance coefficients at different frequency bands, thereby avoiding the significant overhead associated with performing separate measurements for each frequency band.
			
			The AR method proposed in this section directly predicts the unknown covariance matrices without estimating APS, making it relatively straightforward to implement. In the next section, we explore a low-complexity method to acquire APS from the spatial covariance matrix, which is essential for realizing multi-frequency cooperation.

			\section{Maximum Entropy APS Estimation}\label{section4}
			
			

			In this section, we propose a method to estimate the APS with high resolution. We first formulate the estimation problem and analyze it using the ME criterion. Then, we provide the form of the estimated APS and propose a low-complexity algorithm to solve the estimation problem.
			
			For convenience, we define the angular frequencies $({u_i}, {v_i})$ in terms of ${d_i}$, ${\lambda _i}$ and $(u, v)$ as ${{u}_i} = {d_i}u/{\lambda _i}$ and ${{v}_i} = {d_i}v/{\lambda _i}$. Based on Nyquist sampling theory, to avoid spatial aliasing, ${d_i}/{\lambda _i} \le 0.5$ always holds \cite{stoica2005spectral}. Meanwhile, we use a function ${P_k}({u_i},{v_i})$ satisfying ${P_k}({u_i},{v_i}) = {S_k}(u,v)$ for any ${{u}_i} = {d_i}u/{\lambda _i}$ and ${{v}_i} = {d_i}v/{\lambda _i}$ to describe the angular power distribution instead of ${S_k}(u,v)$. With ${P_k}({u_i},{v_i})$, the relationship between the covariance elements and the APS in (\ref{rmn}) can be rewritten as
			\begin{equation}\label{rkmn}
			\scalebox{0.98}{${r_{k,f_i}}(m,n) = \displaystyle\int \!\!\!{\int_{\bf{F}} \!\!{{P_k}({u_i},{v_i}){e^{\bar \jmath 2\pi (m{u_i} + n{v_i})}}d{u_i}d{v_i}} }, (m,n) \in {\bf{\Delta }},$}
			\end{equation}
			where ${\bf{F}}$ represents the range $[ - 0.5,0.5] \times [ - 0.5,0.5]$ and ${\bf{\Delta }} = \{ (m,n):m,n = 0, \pm 1, \cdots , \pm ({N_i} - 1)\}$.

			From (\ref{rkmn}), we observe that the two-dimensional function $r_{k,{f_i}}(m,n)$ and the APS function ${P_k}({u_i},{v_i})$ form a Fourier transform pair. Therefore, if the function $r_{k,{f_i}}(m, n)$ is known for all integers $m$ and $n$, ${P_k}({u_i},{v_i})$ can be expressed as the Fourier transform of $r_{k,{f_i}}(m, n)$, i.e.,
				\begin{equation}\label{Sxi}
				{P_k}({{u_i}},{{v_i}}) = \sum\limits_{m =  - \infty }^\infty  {\sum\limits_{n =  - \infty }^\infty  {{r_{k,{f_i}}(m, n)}{e^{ - {\bar \jmath }2\pi (m{{u_i}} + n{{v_i}})}}} }.
				\end{equation}
				However, for the estimation problem addressed in this paper, only a subset of the values $r_{k,{f_i}}(m, n)$ are known. As a result, there are infinitely many solutions for the APS that satisfy the constraint in (\ref{rkmn}). For example, using the proposed AR model in (\ref{2dARd1ys}), we can derive the corresponding APS through the Fourier transform as
					\begin{equation}\label{Sfar}
					\hat {{P_k}}\left( {{{u_i}},{{v_i}}} \right) = \frac{{{\sigma ^2}}}{{{{\left| {1 + \sum\limits_{q = 1}^{{N_i} - 1} {\sum\limits_{l = 1}^{{N_i} - 1} {{a_1}(q,l){e^{ - {\bar \jmath }2\pi (q{{u_i}} + l{{v_i}})}}} } } \right|}^2}}}.
					\end{equation}	
					Thus, it is necessary to establish suitable criteria and methods to find an APS that not only satisfies the constraint (\ref{rkmn}) but also facilitates efficient multi-frequency transmission.

			\subsection{Problem Formulation}

			In practical propagation environments, channels often exhibit sparsity in the angle domain, with dominant power values concentrated within limited ranges. By leveraging the peak positions in the APS, the locations of scatterers can be determined. To more accurately reflect the power distribution characteristics in the angle domain, the estimated APS requires sharper peaks and a flatter background, i.e., higher resolution. In the field of signal processing, the ME criterion is commonly used to estimate the power spectrum in the frequency domain with the highest resolution \cite{wu2012maximum}. Inspired by this, we apply the ME criterion to APS estimation to obtain results with the highest resolution.
			
			According to \cite{mcclellan1982multidimensional}, the 1D frequency-domain power spectrum can be efficiently estimated using a 1D AR model, which is mathematically equivalent to the spectrum derived from the ME criterion. This equivalence also holds for 1D APS estimation, as demonstrated in \cite{stoica2005spectral}. However, in our scenario, where the 2D APS needs to be estimated, the increased dimensionality introduces significant complexity. Consequently, the equivalence between the AR APS, as shown in (\ref{Sfar}), and the ME APS may no longer hold. To address this, we investigate whether the AR and ME APSs remain equivalent in 2D scenarios.

			\begin{figure*}[t]
				\begin{align}\label{pro1}
			&{{\hat P}_k}\left( {{u_i},{v_i}} \right) = \mathop {\arg \max }\limits_{{P_k}({u_i},{v_i})} \int {\int_{({u_i},{v_i}) \in {\bf{F}}} {{P_k}} } ({u_i},{v_i})d{u_i}d{v_i}, \notag\\
		&{\rm{s}}{\rm{.t}}{\rm{.}}\quad {r_{k,{f_i}}}(m,n) = \int {\int_{({u_i},{v_i}) \in {\bf{F}}} {{P_k}} } ({u_i},{v_i}){e^{\jmath 2\pi (m{u_i} + n{v_i})}}d{u_i}d{v_i},\quad \forall (m,n) \in {\bf{\Delta }}.
			\end{align}
				\hrulefill
			\end{figure*}
			
			From \cite{10118598}, the ME estimation problem can be formulated as (\ref{pro1}), shown at the top of the next page. By applying the Lagrange multiplier method to solve (\ref{pro1}), the ME APS can be expressed in the form of			
			\begin{equation}\label{Sf2d}
				{\hat P_k}({{u_i}},{{v_i}}) = \frac{1}{{C({{u_i}},{{v_i}})}}.
			\end{equation}
			Here, $C({u_i},{v_i})$ is a non-negative polynomial, which can be written as
			\begin{equation}\label{Pf}
				C({u_i},{v_i}) = \sum\limits_{(m,n) \in {\bf{\Delta }}} {c(m,n){e^{ - \bar \jmath 2\pi (m{u_i} + n{v_i})}}},
			\end{equation}
			where $c(m,n)$ is the polynomial coefficient of the term ${{e^{ - {\bar \jmath }2\pi (m{{u_i}} + n{{v_i}})}}}$ satisfying $c(m,n) = {c^*}( - m, - n)$.

			 A comparison of (\ref{Sf2d}) and (\ref{Sfar}) reveals that the key difference lies in the structure of their denominators. Unlike the AR APS in (\ref{Sfar}), the polynomial $C({u_i},{v_i})$ in (\ref{Sf2d}) cannot be factorized into a magnitude-squared polynomial. This structural distinction implies that the 2D ME APS cannot be directly replaced by the AR APS, as is feasible in 1D scenarios. Consequently, a distinct approach beyond the AR method is required to solve (\ref{Sf2d}) for 2D cases. This limitation motivates our adoption of the ME criterion, which provides a more flexible and accurate framework for high-resolution APS estimation in higher-dimensional settings.

			From (\ref{Sf2d}) and (\ref{Pf}), it is intuitive that estimating APS using the maximum entropy criterion  is equivalent to finding the polynomial coefficients $c(m,n)$, such that $ {{\hat P}_k}({{u_i}},{{v_i}})$ in (\ref{Sf2d}) satisfies the constraints in (\ref{rkmn}).  From (\ref{Pf}),  it can be observed that the polynomial $C({{u_i}},{{v_i}})$ can be interpreted as the Fourier transform of a function satisfying
			\begin{equation}\label{ptheta}
				{c_1}(m,n) = \left\{ {\begin{array}{*{20}{c}}
						{c(m,n)},&{(m,n) \in {\bm{\Delta }}},\\
						0,&{{\rm{otherwise}}}.
				\end{array}} \right .
			\end{equation}
			The value of ${{{c_1}}(m,n)}$ is nonzero within the range of ${\bm{\Delta }}$ while being 0 otherwise. Based on (\ref{rkmn}) and (\ref{Sf2d}), the relationship between ${r_{k,{f_i}}(m, n)}$ and the polynomial $C({{u_i}},{{v_i}})$ can then be expressed with ${c_1}(\cdot)$ as
			\begin{align}\label{rkiftft}
				&\quad\quad{r_{k,{f_i}}(m, n)}\notag	\\
				&= \iint_{({u_i},{v_i}) \in {\bf{F}}} {\frac{{{e^{{\bar \jmath }2\pi (m{{u_i}} + n{{v_i}})}}}}{{C({{u_i}},{{v_i}})}}} d{{u_i}}d{{v_i}}\notag	\\
				&= \iint_{({u_i},{v_i}) \in {\bf{F}}} {\frac{{{e^{{\bar \jmath }2\pi (m{{u_i}} + n{{v_i}})}}}}{\sum\limits_{(m',n') \in {\bf{\Delta }}} {c(m',n'){e^{ - \bar \jmath 2\pi (m'{u_i} + n'{v_i})}}} }} d{{u_i}}d{{v_i}}\notag	\\
				&= {{\cal F}^{ - 1}}\left\{ {\frac{1}{{{\cal F}\left\{ {{c_1}(m',n')} \right\}({u_i},{v_i})}}} \right\}(m,n).
			\end{align}
			From the above analysis, the original problem can be transformed into solving for the function ${{{c_1}}(m,n)}$. Since the covariance value ${r_{k,{f_i}}(m, n)}$ and the coefficient ${{{c_1}}(m,n)}$ can be obtained from each other through IF and IFT operations, we propose an iterative algorithm to determine the value that satisfies (\ref{rkiftft}).
			
			\subsection{Solution to the ME Estimation Problem}
			Although the 2D ME estimation problem has been converted to finding the function ${{{c_1}}(\cdot)}$, the nonnegativity of APS and the symmetry of $r_{k,{f_i}}(m, n)$ impose several additional constraints that must be satisfied. According to (\ref{Sf2d}), since the APS with maximum entropy must be nonnegative over the entire angle domain, it follows that
			\begin{equation}\label{tiaojian}
				{\cal F}\left\{ {{c_1}(m,n)} \right\}({u_i},{v_i}) \ge 0,\quad{\cal F}\left\{ {{r_{k,{f_i}}}(m,n)} \right\}({u_i},{v_i}) \ge 0,
			\end{equation}
			always hold for any $({{u_i}},{{v_i}}) \in {\bf{F}}$.  Considering the truncation in (\ref{ptheta}), ${{{c_1}}(m,n)}$ for any integers $m$ and $n$ can be further expressed as
			\begin{equation}
				{c_1}(m,n) = \!\!{{\cal F}^{ - 1}}\!\!\left\{ {\frac{1}{{{\cal F}\left\{ {{r_{k,{f_i}}}(m',n')} \right\}({u_i},{v_i})}}} \right\}(m,n) \cdot w(m,n),
			\end{equation}
			where $w(m,n)$ is a window function satisfying 
			\begin{equation}
				w(m,n) = \left\{ {\begin{array}{*{20}{c}}
						1,&{(m,n) \in {\bm{\Delta }} },\\
						0,&{\rm{otherwise}}.
				\end{array}} \right.
			\end{equation}
			
			
			Next, we employ an iterative method to solve for ${c_1}(m,n)$. We initialize ${c_1}(m,n)$ as $c_1^0(m,n)$, and let $c_1^d(m,n)$ denote the updated coefficients after the $d$-th iteration. Thus, in the $d$-th iteration, a function $r'(m,n)$ can be obtained by substituting $c_1^{d-1}(m,n)$ into (\ref{rkiftft}), i.e.,
			\begin{equation}\label{MEstep1}
				\begin{array}{*{20}{l}}
					{r'(m,n) = {{\cal F}^{ - 1}}\left\{ {\frac{1}{{{\cal F}\left\{ {c_1^{d - 1}(m',n')} \right\}({u_i},{v_i})}}} \right\}(m,n).}
				\end{array}
			\end{equation}
			Once $r'(m,n) = {r_{k,{f_i}}(m, n)}$ holds for all integers $m$ and $n$, the algorithm terminates, and ${c_1^{d - 1}(m,n)}$ is the desired solution for $c_1^{}(m,n)$. However, due to the influence of the size of ${{\bf{R}}_{k,{f_i}}}$, the known range of ${r_{k,{f_i}}(m, n)}$ is limited. To ensure that the constraint in (\ref{rkmn}) always holds, for any $(m,n) \in {\bm{\Delta }}$, $r'(m,n)$ should be replaced by $r_{k,f_i}(m,n)$. Consequently, the final value $r_k^d(m,n)$ after the $d$-th iteration can be expressed as
			\begin{equation}\label{split1}
				\begin{split}
					r_k^d(m,n) &= \left\{ {\begin{array}{*{20}{c}}
							{{r_{k,{f_i}}(m, n)}},\quad{(m,n) \in {\bm{\Delta }}},\\
							{r'(m,n)},\quad{\rm{otherwise}}
					\end{array}} \right.\\
					&= r'(m,n) + [{r_{k,{f_i}}(m, n)} - r'(m,n)]w(m,n).
				\end{split}
			\end{equation}
			Meanwhile, by operating the IFT of (\ref{rkiftft}), a coefficient function $c'(m,n)$ can be calculated by
			\begin{equation}\label{MEstep2}
				c'(m,n) = {{\cal F}^{ - 1}}\left\{ {\frac{1}{{{\cal F}\left\{ {r_k^d(m',n')} \right\}({u_i},{v_i})}}} \right\}(m,n).
			\end{equation}
			Nevertheless, to satisfy (\ref{ptheta}), $c'(m,n)$ should be further corrected to ${c^d_1}(m,n)$ as 
			\begin{equation}
				{c^d_1}(m,n) = \left\{ {\begin{array}{*{20}{c}}
						{c'(m,n)},&{(m,n) \in {\bm{\Delta }}},\\
						0,&{\rm{otherwise}},
				\end{array}} \right.
			\end{equation} 
			i.e., 
			\begin{equation}\label{pmthe}
				c^d_1(m,n) = c'(m,n)w(m,n).
			\end{equation}

		Since $r_k^d(m,n)$ corresponds to the covariance element of a covariance matrix, its Fourier transform should be nonnegative over the range of angular frequencies. However, during the iterative process, due to the substitution of $r'(m,n)$ and the truncation by $w(m,n)$, the nonnegativity constraint in (\ref{tiaojian}) may not always be satisfied. To ensure that $r_k^d(m,n)$ approaches the actual $r_{k,f_i}(m,n)$ while maintaining the nonnegativity of the APS, we introduce a parameter ${\alpha _d} \in [0,1]$ in the $d$-th iteration. This parameter allows a partial substitution of $r'(m,n)$ by $r_{k,f_i}(m,n)$, modifying (\ref{split1}) as follows:
			\begin{equation}
				r_{k}^d(m,n) = \left\{ {\begin{array}{*{20}{c}}
						\!\!\!\!{(1 - {\alpha _d}){r_{k,{f_i}}(m, n)} + {\alpha _d}r'(m,n),\quad\!\!\! (m,n) \in {\bf{\Delta }}}\\
						\!\!\!\!{r'(m,n),\quad \quad \quad \quad \quad \quad \quad \quad\quad \quad {\rm{otherwise}}},
				\end{array}} \right.
			\end{equation}
			i.e.,
			\begin{equation}\label{suanfa1}
				r_{k}^d(m,n) = r'(m,n) + (1 - {\alpha_d} )[{r_{k,{f_i}}(m, n)} - r'(m,n)]w(m,n).
			\end{equation}
			Meanwhile, to guarantee the nonnegativity constraint, as for ${c^d_1}(m,n)$, another parameter ${\beta_d}\in [0,1]$ is introduced to the truncated ${c^d_1}(m,n)$, and (\ref{pmthe}) is modified to 
			\begin{equation}\label{suanfa2}
				{c^d_1}(m,n) = {\beta_d} {c^{d - 1}_1}(m,n) + (1 - {\beta_d} )c'(m,n)w(m,n).
			\end{equation}
			
			\begin{figure*}[ht]
				\normalsize
				\begin{equation}\label{alpham}
				{\alpha _d} = \left\{
				\begin{array}{ll}
				\max \left\{ {\alpha _{d - 1}},1 - {k_d}{\alpha _{\inf }} \right\}, & \mathop {\inf }\limits_{({u_i},{v_i}) \in {\bf{F}}} \left\{ {{\cal F}\left\{ {[{r_{k,{f_i}}}(m,n) - r'(m,n)]w(m,n)} \right\}({u_i},{v_i})} \right\} < 0\vspace{1ex} \\\
					0, & \text{otherwise.}
					\end{array}
					\right.
					\end{equation}
			
			\begin{equation}\label{alpham11}
			{\alpha _{\inf }} = \mathop {\inf }\limits_{({u_i},{v_i}) \in {\bf{F}}_{{u_i},{v_i}}^ - } \left\{ {\frac{{{\cal F}\left\{ {r'(m,n)} \right\}({u_i},{v_i})}}{{\left| {{\cal F}\{ [{r_{k,{f_i}}}(m,n) - r'(m,n)]w(m,n)\} ({u_i},{v_i})} \right|}}} \right\}.
			\end{equation}

			\begin{equation}\label{betam}
					{\beta _d} = \left\{
					\begin{array}{ll}
					\!\!\!(1 - {k_d}) + {k_d}\!\!\mathop {\sup }\limits_{({u_i},{v_i}) \in {\bf{F}}} \left\{ \frac{{\left| {{\cal F}\left\{ {c'(m,n)w(m,n)} \right\}({u_i},{v_i})} \right|}}{{\left| {{\cal F}\left\{ {c'(m,n)w(m,n)} \right\}({u_i},{v_i})} \right| + \left| {{\cal F}\left\{ {c_1^{d - 1}(m,n)} \right\}({u_i},{v_i})} \right|}} \right\}, &\!\!\!\!\! \mathop {\inf }\limits_{({u_i},{v_i}) \in {\bf{F}}} \left\{ {{\cal F}\left\{ {c'(m,n)w(m,n)} \right\}({u_i},{v_i})} \right\} < 0, \\
					0, & \text{otherwise}.
					\end{array}
					\right.
				\end{equation}
				\hrulefill
			\end{figure*}
			When (\ref{rkmn}) is satisfied, we have ${\alpha _d} = {\beta_d} = 0$. Otherwise, the algorithm will terminate when ${\alpha _d} = {\beta_d} = 1$, as ${r_{k,{f_i}}(m, n)}$ will no longer affect $r_{k}^d(m,n)$ in this case. Therefore, during the iteration, it is necessary to find the minimum ${\alpha _d}$ and ${\beta_d}$ that are closest to 0 in order to make $r_{k}^d(m,n)$ approach the actual ${r_{k,{f_i}}(m, n)}$, while ensuring that (\ref{tiaojian}) is satisfied. The minimum ${\alpha _d}$ under the constraints in (\ref{tiaojian}) is given by (\ref{alpham}), where ${{\alpha _{\inf }}}$ is defined in (\ref{alpham11}), and the minimum ${\beta_d}$ is provided in (\ref{betam}) at the top of this page, where ${\bf{F}}_{{{u_i}},{{v_i}}}^ - $ represents the range of angular frequencies defined as
			\begin{equation}\label{Jxi}
				\begin{array}{l}
					\vspace{1ex} {\bf{F}}_{{{u_i}},{{v_i}}}^ -  = \{ ({{u_i}},{{v_i}}):\\
					\quad\quad\quad\quad{{\cal{F}}}\{ [{r_{k,{f_i}}(m, n)} - r'(m,n)]w(m,n)\} ({{u_i}},{{v_i}}) < 0\} 
				\end{array}
			\end{equation}
			and ${k_d}$ is the convergence rate factor of the $d$th iteration. The justification for the design choices of ${\alpha_d}$ and ${\beta_d}$ is provided in Appendix B.

			We define
			\begin{equation}\label{vard}
				{\varepsilon _d} = \sum\limits_{(m,n) \in {\bf{\Delta }}} {\frac{{|{r_{k,{f_i}}}(m,n) - r_k^{d - 1}(m,n){|^2}}}{{|{r_{k,{f_i}}}(m,n){|^2}}}} 
			\end{equation}
			to measure the estimated error.
			Ideally, when the algorithm terminates, $\varepsilon_d = 0$. However, to maintain the nonnegativity of the power spectrum during the actual implementation of the algorithm, there will always be some error between $r_k^{d-1}(m,n)$ and the actual $r_{k,f_i}(m,n)$. Therefore, the algorithm threshold is set as ${\varepsilon _{\min }} \ne 0$. In this case, the iteration terminates when ${\varepsilon_d} \leq {\varepsilon _{\min }}$.
			
			\begin{algorithm}[h]
				\caption{ME APS Estimation} 
				\label{algorithm}
				\begin{algorithmic}[1]
					\footnotesize
					\Require
					$r_k^0(m,n)$, ${c^0_1}(m,n)$, ${\alpha _{0}}=0$, ${\beta _{0}}=0$, ${k_0}=1$, ${\varepsilon_0} = 1$, ${\varepsilon _{\min }}$, and the length of FFT;
					\Ensure
					${P_k}({{u_i}},{{v_i}})$;
					\State Initialize $r_k^0(m,n) = {r_k}(0,0)\delta (m,n)$, ${c^0_1}(m,n) = \frac{1}{{{r_k}(0,0)}}\delta (m,n)$, ${\alpha _0} = {\beta _0} = 0$, $d = 0$;
					\While {${\varepsilon_d} > {\varepsilon _{\min }}$}
					\State $d \leftarrow d + 1$;
					\State Calculate $\varepsilon_d$ based on (\ref{vard});
					\If{${\varepsilon _d} > {\varepsilon _{d - 1}}$} 
					\State 
					${k_d} = 0.5{k_{d - 1}}$;
					\EndIf 
					\State $r'(m,n) = {{\cal F}^{ - 1}}\left\{ {\frac{1}{{{\cal F}\left\{ {c_1^{d - 1}(m',n')} \right\}({u_i},{v_i})}}} \right\}(m,n)$;
					\State Calculate ${\alpha _d}$ and ${\beta_d}$ based on (\ref{alpham}) and (\ref{betam});
					\State Correct $r'(m,n)$ with $r_k^{}(m,n)$ for all ${(m,n)} \in {\bm{\Delta }}$ to obtain $r_{k}^d(m,n)$ based on (\ref{suanfa1}) ;
					\State Calculate $\varepsilon_d$ based on (\ref{vard});
					\State $c'(m,n) = {{\cal F}^{ - 1}}\left\{ {\frac{1}{{{{\cal F}^{ - 1}}\left\{ {r_k^d(m',n')} \right\}({u_i},{v_i})}}} \right\}(m,n)$;
					\State Correct $c'(m,n)$ based on (\ref{suanfa2}) to obtain ${c^d_1}(m,n)$ ;
					\State Substitude ${c^d_1}(m,n)$ corresponding to ${(m,n)} \in {\bf{\Delta }}$ into (\ref{Sf2d}) and (\ref{Pf}) to obtain the estimated spectrum.

					\EndWhile
				\end{algorithmic}
			\end{algorithm}

			In actual calculations, the FT and IFT operations in the algorithm are replaced by 2D FFT and IFFT, respectively. The continuous range $[-0.5, 0.5]$ is then discretized into $B$ values, where $B$ represents the FFT size during the iteration. To prevent aliasing and account for the size of ${N_i}$, it is crucial to select an appropriate value for $B$ throughout the iterative process. As for the convergence rate factor, we set ${k_0} = 1$ and update ${k_d} = 0.5{k_{d - 1}}$ in the $d$th iteration whenever ${\varepsilon _d} > {\varepsilon _{d - 1}}$ to avoid divergence. Additionally, the selection of the initial value must satisfy the conditions specified in (\ref{rkmn}). To facilitate the iteration, the Dirac delta function is often chosen as the initial value for the function $r_k^0(m,n)$, while its reciprocal is set as the initial value for $c^0_1(m,n)$. Algorithm \ref{algorithm} provides a detailed procedure for the method, following which an estimated value of the ME APS can be obtained.%

			

			Through the above theoretical derivation, in a multi-frequency system, the ME estimation can derive the APS that satisfies the maximum entropy condition. Given that the APSs corresponding to all frequency bands can be approximately regarded as similar, we can leverage the APS to achieve cooperative transmission across different frequency bands. For instance, in a dual-frequency system comprising two frequencies ${f_1}$ and ${f_2}$, where ${f_1} < {f_2}$, the APS can be estimated from ${{\bf{R}}_{k,{f_1}}}$ and then directly utilized to facilitate robust transmission at ${f_2}$. This multi-frequency cooperation eliminates the need for training and measurements associated with uplink detection at ${f_2}$, thereby significantly reducing the system overhead.



			\section{Simulation Results}\label{section5}
			In this section, simulation results are presented to validate the effectiveness of the proposed approaches. 
			
			We adopt two types of channel modeling approaches in this section. The first approach follows the channel modeling method described in \cite{lu2019robust,mine}, where the APS is initially generated, and the spatial domain channel is subsequently constructed using the APS. Since the real APS is known in this modeling approach, it facilitates the evaluation of the proposed APS estimation method. The second approach utilizes the quasi-deterministic radio channel generator (QuaDRiGa) \cite{jaeckel2014quadriga}, a widely-used channel model capable of accurately simulating real-world channels across various environments. This channel model is employed to validate the effectiveness of multi-frequency cooperative transmission.

			\subsection{Performance of Covariance Matrix Prediction}

			We begin by evaluating the performance of the covariance prediction method discussed in Section~\ref{section3}.
			
			We model a dual-frequency system comprising the frequencies $f_1$ and $f_2$, where ${f_1} < {f_2}$, using the channel modeling method described in \cite{lu2019robust} and \cite{mine}. In practical applications, since the continuous APS is challenging to handle, the angular domain is typically discretized for easier analysis. Consequently, the APS to be estimated is also represented in a discretized form as ${S_k}(u,v)$. The discretization process involves dividing the range $[-1,1]$ for both $u$ and $v$ into equal intervals. Let $B$ denote the number of grid points, and let $\{ {u_1},{u_2}, \cdots ,{u_B}\} $ and $\{ {v_1},{v_2}, \cdots ,{v_B}\} $ represent the sets of discrete directional cosines in the two dimensions. Then, ${{\bf{R}}_{k,{f_i}}}$ in (\ref{Rk}) can be expressed as
			\begin{equation}\label{RKK}
			{{\bf{R}}_{k,{f_i}}} = \sum\limits_{b = 1}^B {\sum\limits_{b' = 1}^B {{{\hat S}_k}({u_b},{v_{b'}}){{\bf{v}}_{{f_i}}}\left( {{u_b},{v_{b'}}} \right){\bf{v}}_{{f_i}}^{\rm{H}}\left( {{u_b},{v_{b'}}} \right)} },
			\end{equation}
			where $\hat{S}_k(u_b, v_{b'})$ is the power in the direction  $({u_b},{v_{b'}})$.

			To generate $\hat{S}_k(u_b, v_{b'})$, we selected $P$ directions corresponding to the central directions of incident paths, where each path represents a spatial cluster. Subsequently, we implemented the methodology proposed in \cite{9786750} to calculate the power distribution across different clusters in each direction. Given the assumption that ${d_i}/{\lambda_i} = 0.5$ for $i = 1,2$, and utilizing ${\bf{R}}_{k,{f_1}}^{}$ as the initial condition, the high-dimensional ${\bf{R}}_{k,{f_2}}^{}$ can be effectively predicted.




			We evaluate the performance of the proposed AR prediction method by comparing it with the conventional linear extrapolation technique presented in \cite{ali2016estimating}. For this comparative analysis, we fix $N_1 = 8$ and investigate the influence of array dimensions by considering two configurations: ${N_2} = 10$ and ${N_2} = 12$, respectively.
			
\begin{figure}[htbp]
				\centering
				\includegraphics[scale = 0.5]{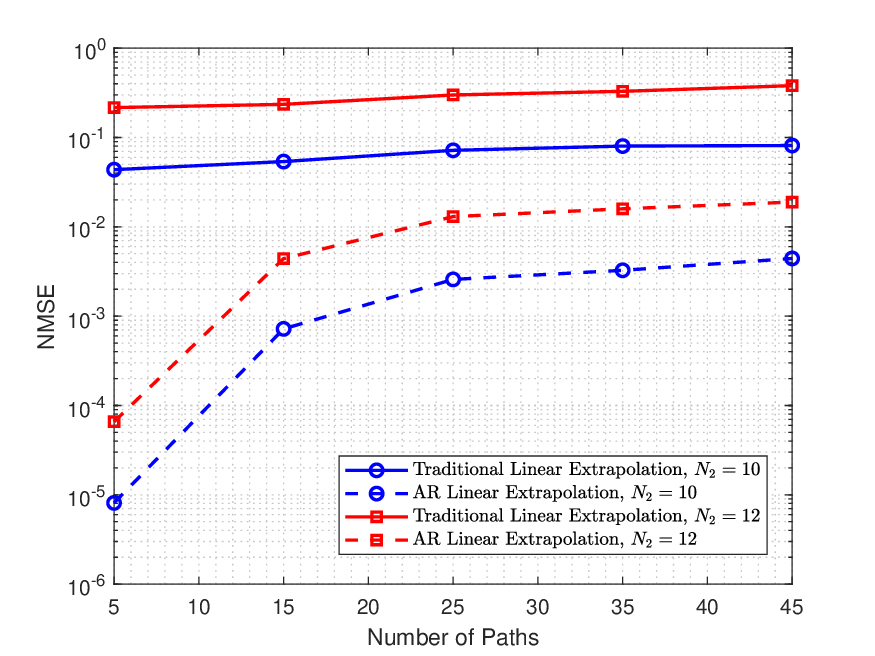}
				\caption{NMSE of different covariance matrix prediction methods versus the number of paths, ${N_1} =8$, ${N_2}= 10, 12$.}
				\label{2DAR}
			\end{figure}

			To quantify the accuracy of covariance matrix prediction, we employ a performance metric based on the normalized mean square error (NMSE), defined as
			\begin{equation}\label{MSE}
				{\rm{NMSE}} = \frac{{||{{{\bf{\hat R}}}_{k,{f_2}}} - {{\bf{R}}_{k,{f_2}}}||_F^2}}{{||{{\bf{R}}_{k,{f_2}}}||_F^2}},
			\end{equation}
	    	where a lower NMSE value corresponds to better prediction accuracy. Fig. \ref{2DAR} illustrates the NMSE performance of both covariance matrix prediction methods as the number of paths $P$ varies, with the system's signal-to-noise ratio (SNR) maintained at 30 dB throughout the simulation.

			


		As demonstrated in Fig. \ref{2DAR}, the proposed AR method consistently outperforms the traditional approach by effectively leveraging the information from ${{\bf{R}}_{k,{f_1}}}$. Furthermore, a comparison between the results for ${N_2} = 10$ and ${N_2} = 12$ reveals an increase in NMSE for the latter case, primarily due to the larger number of unknown elements to be estimated. The NMSE performance of the proposed AR method is closely related to the number of paths, a consequence of the linearity inherent in the AR model as expressed in (\ref{2dARd1}). From (\ref{RKK}), ${r_{k,f_i}}(m,n)$ can be interpreted as the superposition of contributions from all paths in the spatial domain. This implies that estimating ${r_{k,f_i}}(m,n)$ is equivalent to resolving these paths. However, the AR model relies on solving a system of equations, as shown in (\ref{Rb=1}), and the number of unknowns that can be resolved by such a system is inherently limited. When the number of paths exceeds the number of parameters that can be solved by the equation set, the AR method fails to distinguish all paths, resulting in performance degradation. Consequently, the NMSE performance of the AR method deteriorates as $P$ increases. 

			\subsection{Performance of APS Estimation}

		The aforementioned covariance prediction method omits the estimation of APS. However, in certain scenarios, APS estimation is indispensable. In this subsection, we evaluate the effectiveness of the proposed APS estimation method. To demonstrate the resolution enhancement achieved by the ME criterion, we compare the APS derived from the proposed Algorithm~\ref{algorithm} with the AR spectrum provided in (\ref{Sfar}).

		\begin{figure}[!t]
			\centering	
			\subfloat[]{\includegraphics[width=1.8in]{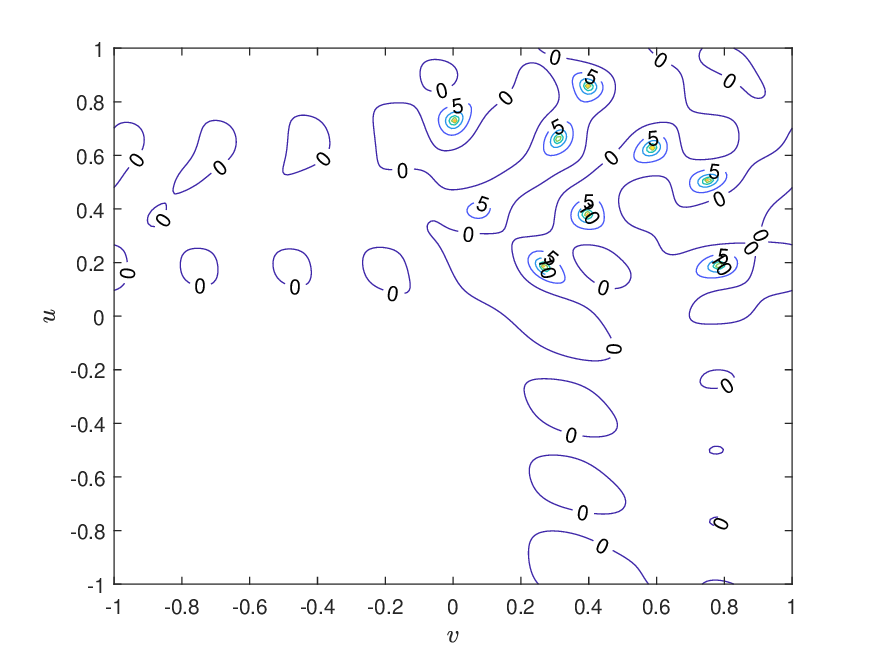}%
				\label{fig_first_case1_10}}
			\hspace{-5mm}
			\hfil
			\subfloat[]{\includegraphics[width=1.8in]{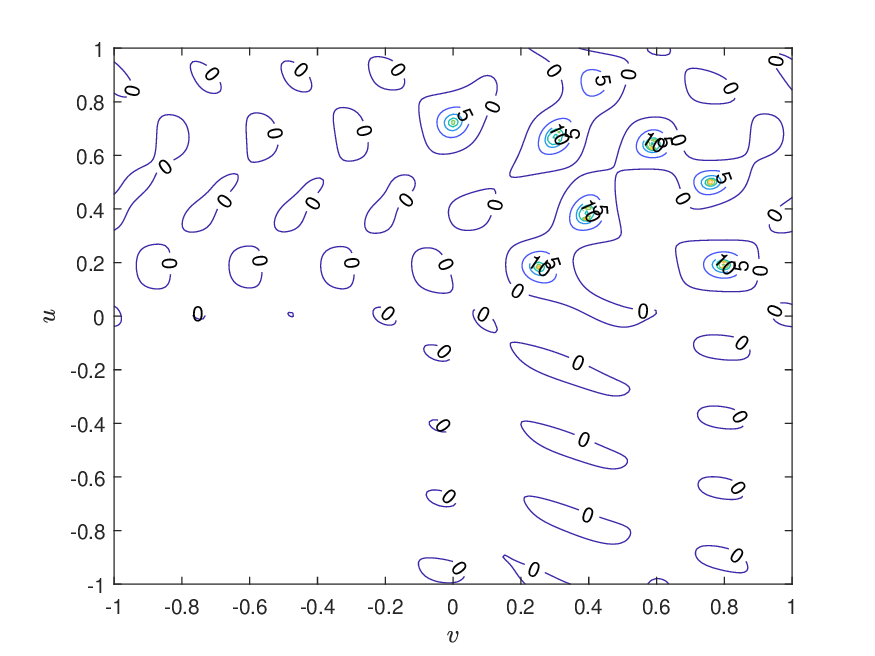}%
				\label{fig_second_case2_10}}
			\caption{APS estimated by different methods, ${N_i} = 8$, $P=8$. (a) APS estimated by AR method; (b) APS estimated by ME method.} 
			\label{fig_sim_aps10}
			
		\end{figure}

			
			We first consider a UPA scenario with a relatively small number of paths, i.e., $P=8$. The scenario is constructed based on the model in (\ref{RKK}), and we utilize an $8 \times 8$ UPA. The powers $\{\gamma_p\}$ of different paths are generated according to the distribution described in \cite{4277071}. Assuming an SNR of $10$ dB, Fig.~\ref{fig_sim_aps10} illustrates the APSs obtained by the AR method and the ME method, respectively. In this case, both methods successfully produce APS estimation results with 8 distinct peaks. By analyzing the peak positions in the APS image, the locations of clusters in the propagation environment can be accurately identified.
			\begin{figure}[!t]
				\centering
				\subfloat[]{\includegraphics[width=1.8in]{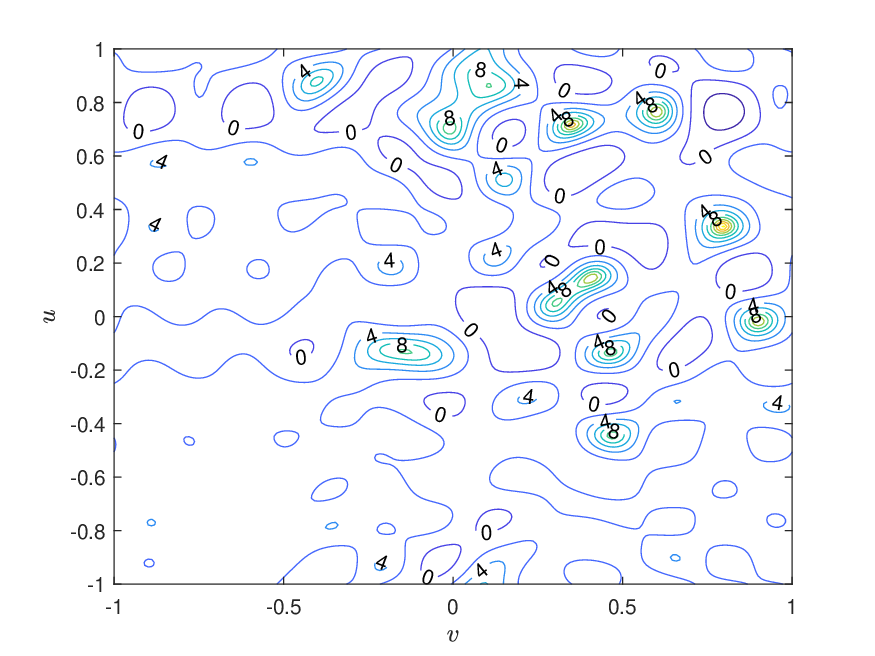}%
					\label{fig_first_case1}}
				\hspace{-5mm}
				\hfil
				\subfloat[]{\includegraphics[width=1.8in]{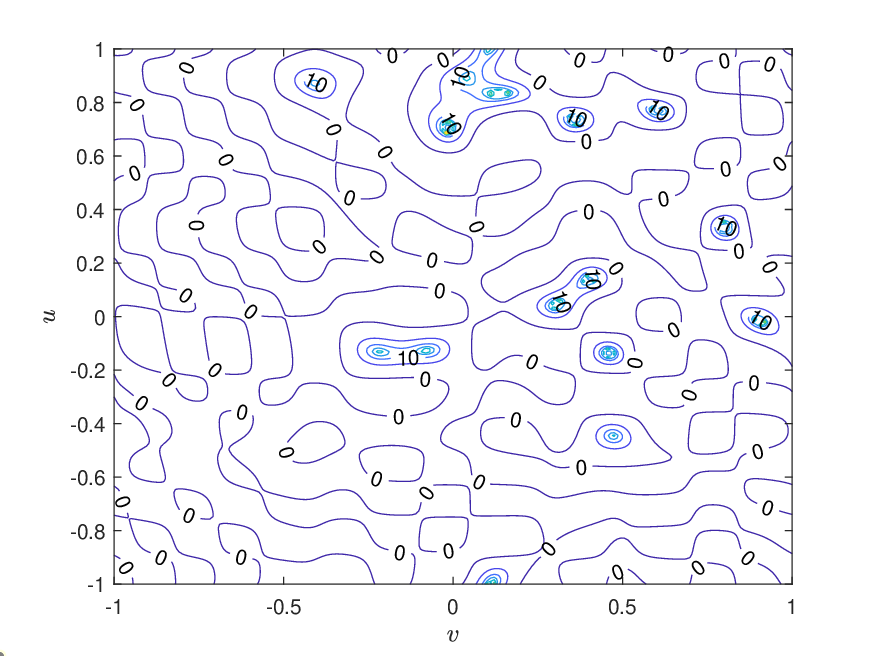}%
					\label{fig_second_case2}}
				\caption{APS estimated by different methods, ${N_i} = 8$, $P=15$. (a) APS estimated by AR method; (b) APS estimated by ME method.} 
				\label{fig_sim3}
			\end{figure}
			To further evaluate the performance of the ME estimation, we increase the number of paths to $P = 15$, and the corresponding results are presented in Fig.~\ref{fig_sim3}. As the number of paths increases, the ME APS demonstrates significantly higher resolution compared to the AR APS. Notably, even in scenarios with closely spaced paths, the ME APS effectively resolves them. Consequently, all directions of the different paths can be precisely determined from the 15 peaks of the ME APS. Therefore, as the number of paths in the system increases, the ME method achieves higher-resolution estimation results compared to the AR method.


			Fig.~\ref{fig_sim_aps10} and Fig.~\ref{fig_sim3} compare the performance of the two APS estimation methods in terms of resolution. To further investigate the impact of system parameters on their performance, we evaluate their behavior with varying values of $N_i$ and $P$. For ease of analysis, the powers of different directions can be represented as a $B \times B$ dimensional matrix ${{\bf{\Omega }}_k}$, where the $(b,b')$-th element is $\tilde{S}_k(u_b, v_{b'})$. Correspondingly, the estimated APS can also be expressed as a matrix, denoted as ${{{\bf{\hat \Omega }}}_k}$. To quantify the APS estimation performance, we compute the NMSE between ${{{\bf{\hat \Omega }}}_k}$ and ${{\bf{\Omega }}_k}$ under different parameter configurations.

\begin{figure}[!t]
		   	\centering	
		   	\subfloat[]{\includegraphics[width=1.8in]{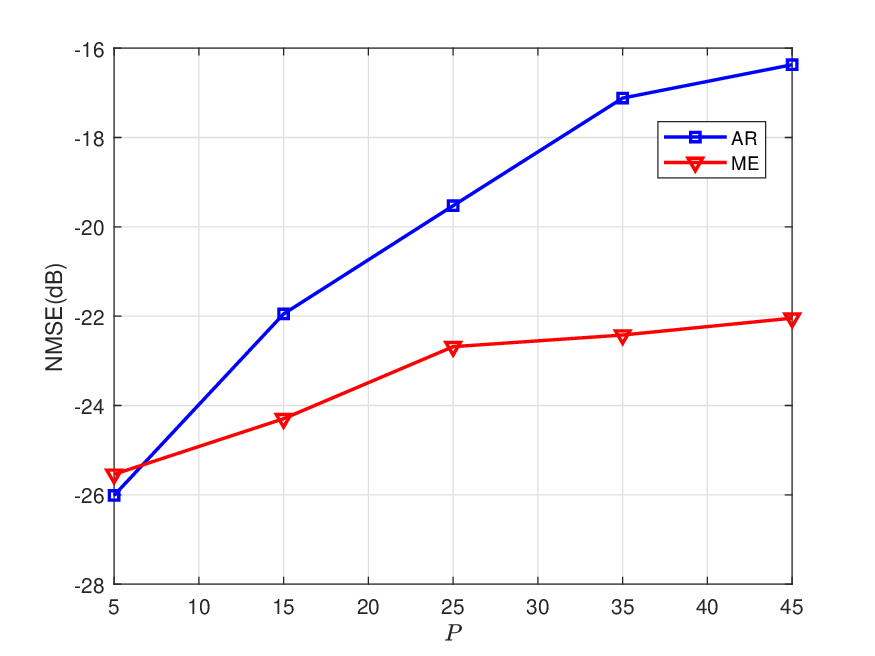}%
		   		\label{KLP}}
		   	\hspace{-5mm}
		   	\hfil
		   	\subfloat[]{\includegraphics[width=1.8in]{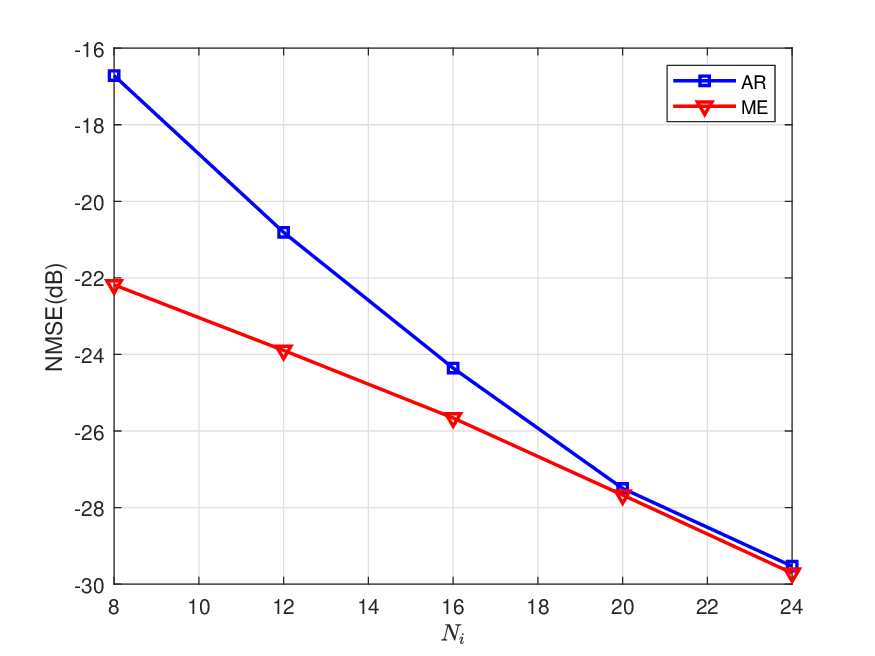}%
		   		\label{KLN}}
		   	\caption{APS estimation performance under different parameter settings. (a) APS estimation performance versus $P$, ${N_i} = 8$; (b) APS estimation performance versus $N_i$, $P = 45$.}
		   	\label{KL}
		   	
		   \end{figure}

		   Fig.~\ref{KL} illustrates the NMSE performance of APS estimation under varying parameter settings. Fig.~\ref{KL}(a) reveals that the performance of the AR method deteriorates as the number of paths, $P$, increases. In contrast, the ME method demonstrates greater robustness to variations in the number of paths. Fig.~\ref{KL}(b) depicts the APS estimation performance with a varying number of antennas. It is evident that the performance of both methods improves with increasing $N_i$, as a larger number of antennas enhances the spatial resolution, thereby enabling more accurate APS estimation.
		   
			\subsection{Performance of Multi-frequency Cooperative Transmission}
		
			To further validate the accuracy of the estimated APS and apply the estimated APS to multi-frequency cooperative transmission, we extend the theoretical multi-frequency model in (\ref{Rk}) to more complex scenarios using QuaDRiGa \cite{jaeckel2014quadriga}. In the context of downlink transmission in massive MIMO systems, the sum rate is a critical metric for assessing transmission quality. For a communication system with $K$ users, the transmission rate of the $k$th user can be expressed as
			\begin{align}
			{R_{k,{f_i}}} = &{\mathbb E}\left\{ {{{\log }_2}\left( {{w_{k,{f_i}}} + {\bf{g}}_{k,{f_i}}^{\rm{H}}{{\bf{p}}_{k,{f_i}}}{\bf{p}}_{k,{f_i}}^{\rm{H}}{\bf{g}}_{k,{f_i}}^{}} \right)} \right.\left. { - {{\log }_2} {{{w}}_{k,{f_i}}}} \right\},
			\end{align}
			where ${{{\bf{p}}_{k,{f_i}}}}$ is the precoding vector and 
			\begin{equation}
			{w_{k,{f_i}}} = \sum\limits_{l \ne k} {\mathbb E}{\left\{ {{\bf{g}}_{k,{f_i}}^{\rm{H}}{{\bf{p}}_{l,{f_i}}}{\bf{p}}_{l,{f_i}}^{\rm{H}}{\bf{g}}_{k,{f_i}}^{}} \right\}}  + {\sigma ^2}.
			\end{equation}
			Therefore, the sum rate of all the users at $f_i$ can be calculated as 
			\begin{equation}\label{rate}
				{R_{{\rm{sum}},{f_i}}} = \sum\limits_{k = 1}^K {{R_{k,{f_i}}}}.
			\end{equation}

			\newcolumntype{L}{>{\hspace*{-\tabcolsep}}l}
			\newcolumntype{R}{c<{\hspace*{-\tabcolsep}}}
			\definecolor{lightblue}{rgb}{0.93,0.95,1.0}
			\begin{table}[!t]
				\renewcommand\arraystretch{1.5}
				\footnotesize
				\caption{Simulation parameters}	\label{last}
				\centering
				\setlength{\tabcolsep}{1.1mm}{} 
				\begin{tabular}{LcRcR}
					\toprule
					Parameter & & Value\\
					\midrule
					\rowcolor{lightblue}
					Array Size && ${8 \times 8}$, ${10 \times 10}$, ${12 \times 12}$&   \\  
					Inter-antenna spacing && $0.5$&\\
					\rowcolor{lightblue}
					Carrier frequency ${f_1}$, ${f_2}$, ${f_3}$ && 2.4, 3.0, 3.6 (GHz)&\\
					SNR && -20 $ -  $20 (dB)	
					\\
					\rowcolor{lightblue}
					Scenario &&  LOS, NLOS&\\    
					Number of users $K$&& 12&\\	
					\rowcolor{lightblue}		
					User speed &&20 (km/h)&\\
					Number of slots&& 200&\\
					\rowcolor{lightblue}		
					Number of blocks in each slot&&5&\\		
					\bottomrule
				\end{tabular}
			\end{table}

			Some typical precoding methods are implemented to design ${{{\bf{p}}_{k,{f_i}}}}$ based on instantaneous CSI, while \cite{lu2019robust} and \cite{you2015pilot} improved previous methods using statistical CSI. However, in practical communication systems, sometimes limited by hardware constraints, channels of some frequency bands lack uplink detection signals, and CSI is not directly available. In such cases, after the user terminal estimates the downlink instantaneous CSI, the BS acquires the instantaneous CSI through feedback from users. The feedback mechanism involves the user comparing the estimated CSI with a collection of codebooks and feeding back the codebook with the highest correlation \cite{kaltenberger2008performance}, which is referred to as the precoding matrix indicator (PMI) \cite{schwarz2010mutual}. The feedback codebook can be regarded as a quantized instantaneous CSI, enabling the matched filter (MF) method and the regularized zero forcing (RZF) method to be performed. As for statistical CSI, due to the lack of uplink detection, the long-term information cannot be obtained from measurement or feedback but can be transformed from that of other bands. According to \cite{lu2019robust}, the discrete APS can then be used to generate a precoding vector, enabling robust transmission under nonideal conditions. To assess the feasibility of multi-frequency robust transmission, we employ various methods to estimate the APS and subsequently apply the estimated APS to the transmission process.
			
			Simulations are conducted in a triple-frequency system comprising channels at ${f_1} = 2.4$ GHz, ${f_2} = 3.0$ GHz, and ${f_3} = 3.6$ GHz. The BS is equipped with UPAs of varying sizes, while each user is equipped with a single antenna. Detailed parameter settings are presented in Table \ref{last}. The time resource is divided into multiple time slots, each of which consists of several blocks. The first block in each slot is utilized for detection, and the subsequent blocks are dedicated to transmissions. Consequently, the instantaneous CSI used for MF and RZF in each slot corresponds to the quantized CSI from the first block, leading to outdated CSI. To address this issue and ensure robust transmission at ${f_2}$ and ${f_3}$, we extract APS from the spatial covariance matrix at ${f_1}$.
			
					\begin{figure}[!t]
				\centering
				\subfloat[Sum rate performances at ${f_2}$ with ${N_2} = 10$. ]{\includegraphics[scale = 0.5]{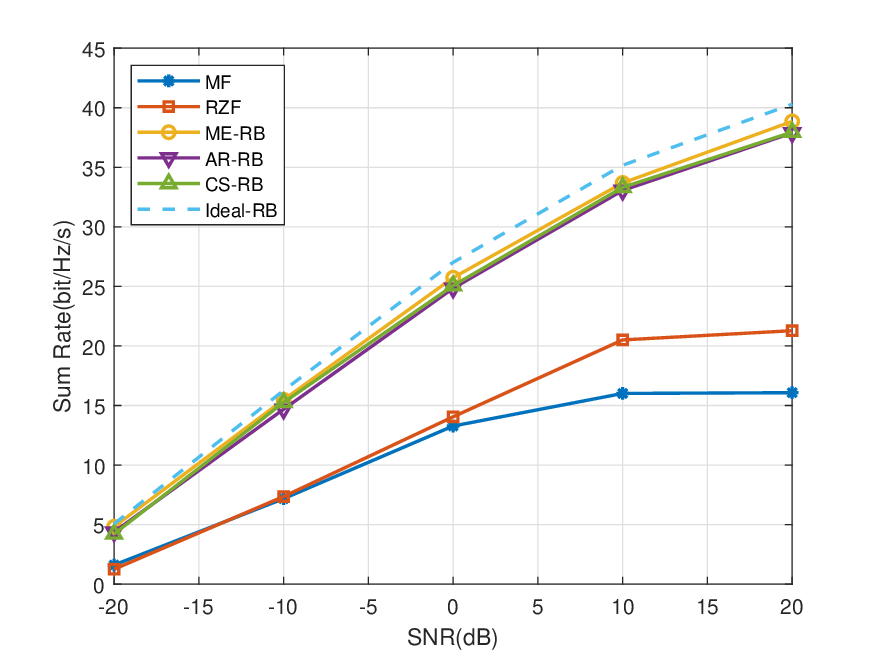}%
					\label{fig_first_case3}}
				\\
				\subfloat[Sum rate performances at ${f_3}$ with ${N_3} = 12$.]{\includegraphics[scale = 0.5]{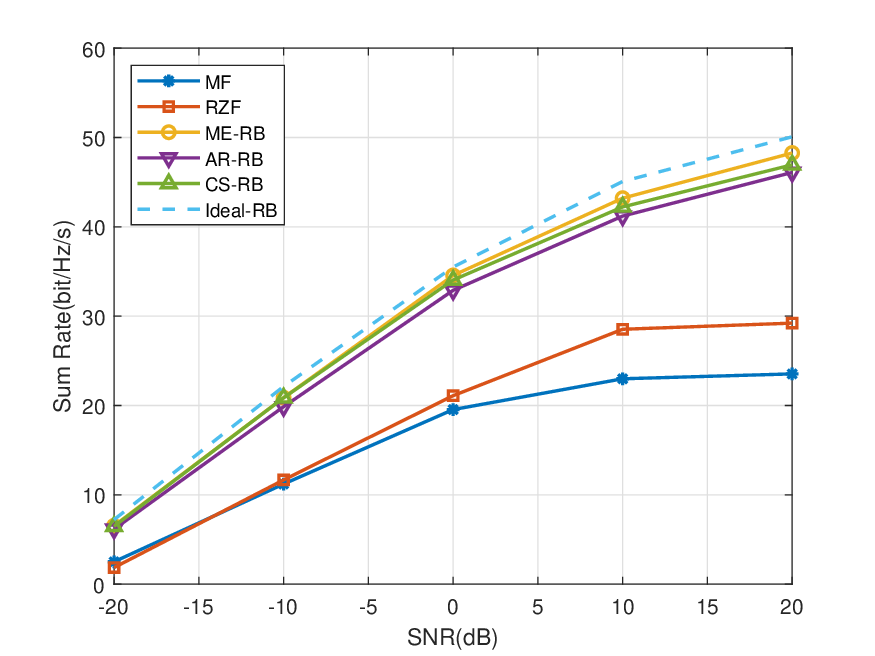}%
					\label{fig_second_case3}}
				\caption{Sum rate performances of different precoding methods vesus SNR in LOS scenario.}
				\label{fig_sim5}
			\end{figure}

		   During the uplink detection at ${f_1}$, ${{\bf{ R}}_{k,{f_1}}}$ can be acquired from the measurement data with the method in \cite{10146318}. The discrete APS can subsequently be estimated from ${{\bf{\hat R}}_{k,{f_1}}}$ using Algorithm \ref{algorithm}. In practical scenarios, we use 2D FFT and IFFT instead of the FT and IFT operations in Algorithm \ref{algorithm}, with the FFT size $B$ set to 32. In this case, the continuous ranges of $u$ and $v$ in ${S_k}\left( {u,v} \right)$, i.e., $[-1,1]$, are quantized into 32 grid points, and the estimated discrete APS can be regarded as a $32 \times 32$ sample of ${S_k}\left( {u,v} \right)$. To further compare the performance of different APS estimation methods, in addition to the ME estimation, we also employ the CS-based method in \cite{ali2019spatial} and the AR method to estimate the APS. The APS estimated by all the aforementioned methods is then used in the downlink robust precoding design for $f_2$ and $f_3$, and their performance is evaluated in terms of the sum rate.

			Simulations are implemented in the light-of-sight (LOS) scenario and non light-of-sight (NLOS) scenario, respectively. We first test the sum rate performance of different precoding methods in a LOS scenario, and the results of different precoding methods are shown in Fig. \ref{fig_sim5}, including:
			\begin{itemize}
				\item[$\bullet$]
				\emph{MF}: ME precoding with the PMI at the current frequency band;
				
				\item[$\bullet$]
				\emph{RZF}: RZF precoding with the PMI at the current frequency band;

				\item[$\bullet$]
				\emph{ME-RB}: Robust precoding with the PMI at the current frequency band and the APS estimated at $f_1$. The APS is estimated by the ME method;
				
				\item[$\bullet$]
				\emph{AR-RB}: Robust precoding with the PMI at the current frequency band and the APS estimated at $f_1$. The APS is estimated by the AR method in (\ref{Sfar});
				
	        	\item[$\bullet$]
		        \emph{CS-RB}:  Robust precoding with the PMI at the current frequency band and the APS estimated at $f_1$. The APS is estimated by the CS-based method \cite{ali2019spatial};
		        
		        \item[$\bullet$]
		        \emph{Ideal-RB}:  Robust precoding with the PMI at the current frequency band and the ideal APS, i.e., the APS acquired at the current frequency band.
			\end{itemize}

        	Fig. \ref{fig_sim5} compares the performance of different precoding methods at ${f_2}$ and ${f_3}$. Since both the CS-based method and the ME method estimate the APS through iterative processes, we set the number of iterations to $100$ for these two methods during APS estimation. The results demonstrate that the robust transmission method, which utilizes the APS estimated from ${f_1}$, significantly outperforms traditional methods such as MF and RZF. In mobile environments, MF and RZF design precoding matrices based on instantaneous CSI, which fails to address outdated CSI or suppress user interference effectively. In contrast, the robust precoding method mitigates these issues by leveraging statistical CSI, leading to substantial performance improvements. This enhancement underscores the feasibility and effectiveness of multi-frequency robust transmission. Notably, in this scenario, all three multi-frequency cooperative precoding algorithms achieve comparable performance. The AR-RB and ME-RB algorithms exhibit similar performance due to the limited number of paths in the LOS scenario. This observation aligns with the results in Fig. \ref{fig_sim_aps10}, where both the AR and ME methods achieve high-resolution APS estimation when the number of paths is small.

			 Although the sum rate performances of ME-RB, AR-RB and CS-RB are similar in Fig. \ref{fig_sim5}, the  computational complexities of these three methods are quite different. Both CS-RB and ME-RB methods estimate the APS at $f_1$ through iterative processes. According to \cite{ali2019spatial}, after $D$ iterations, the computational complexity of the CS-based APS estimation  is ${\cal O}\left( {\sum\limits_{d = 1}^D {\left( {N_1^2{B^2} + dN_1^2 + {d^3}} \right)} } \right)$. Meanwhile,  due to its implementation with 2D FFT operations, the complexity of the proposed ME method is ${\cal O}\left( {D{B^2}{\log_2} B} \right)$. Unlike the other two methods, the AR method does not use an iterative process but directly computes the APS.  From its procedures in (\ref{proc1})-(\ref{vi}), the corresponding computational complexity of the AR method is ${\cal O}\left( {|{{\overline {\cal S} }_{{\rm{AR}}}}{|^3}} \right)$. In our setting with ${N_1} =8$, $B = 32$, and $D = 100$, the AR method exhibits the lowest computational complexity, while the CS-based method has the highest computational complexity. Therefore, in LOS scenarios, the AR method is a more efficient choice for estimating the APS to enable multi-frequency cooperative transmission.

					\begin{figure}[!t]
				\centering
				\subfloat[Sum rate performances at ${f_2}$ with ${N_2} = 10$. ]{\includegraphics[scale = 0.5]{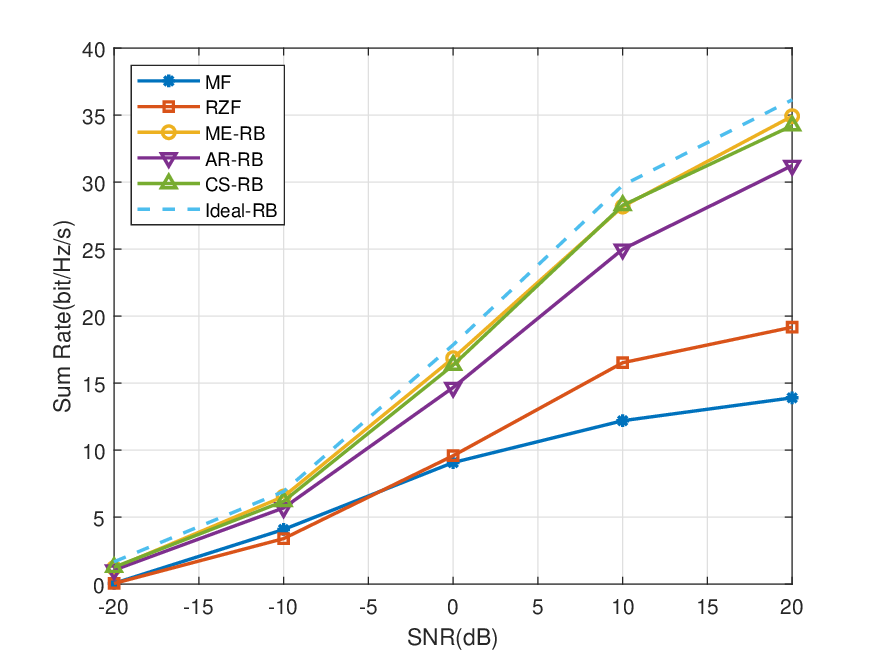}%
					\label{fig_first_case4}}
				\\
				\subfloat[Sum rate performances at ${f_3}$ with ${N_3} = 12$.]{\includegraphics[scale = 0.5]{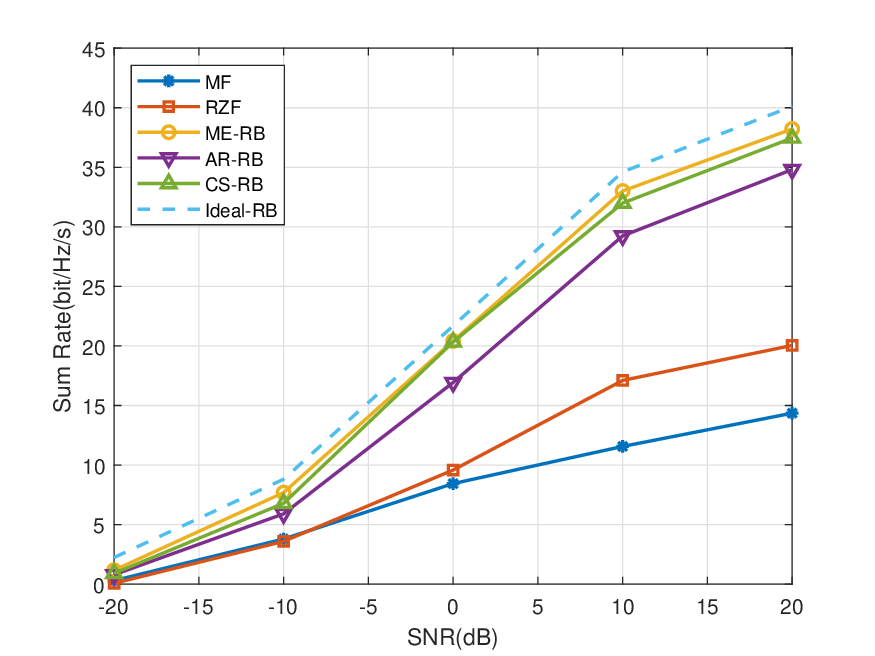}%
					\label{fig_second_case4}}
				\caption{Sum rate performances of different precoding methods vesus SNR in NLOS scenario.}
				\label{fig_sim4}
			\end{figure}
			
			We further evaluate the performance of multi-frequency cooperation in the NLOS scenario. The simulation results, as illustrated in Fig. \ref{fig_sim4}, demonstrate that the robust precoding method exhibits a significant performance gain compared to RZF and MF. Moreover, the propagation scenario also influences the performance of the proposed method. Notably, the number of propagation paths increases substantially in NLOS scenarios compared to LOS scenarios. Consequently, CS-RB and ME-RB outperform AR-RB, which is consistent with the observations in Fig. \ref{fig_sim_aps10} and Fig. \ref{fig_sim3}. These results suggest that as the number of paths increases, the ME method achieves superior APS estimation accuracy compared to the AR method. Additionally, the high-resolution characteristics of the ME estimation enable it to better capture the statistical features of the channel in the angle domain as the number of paths grows. Although both CS-RB and ME-RB employ iterative methods to solve the APS, their distinct optimization objectives lead to subtle differences in performance. Given that the ME method demonstrates lower computational complexity per iteration, ME-RB proves to be more suitable than CS-RB for achieving robust multi-frequency transmission in NLOS scenarios.

			Considering both implementation complexity and channel transmission performance, the AR method is well-suited for extracting statistical CSI to support multi-frequency robust transmission in LOS scenarios. Meanwhile, for NLOS scenarios, the ME algorithm is more appropriate for acquiring statistical CSI to ensure multi-frequency robust transmission.
			
			To evaluate the performance of the proposed multi-frequency cooperative transmission method in higher-mobility scenarios, we adjust the user speeds to 60 km/h and 80 km/h based on the parameters defined in Table \ref{last}. For the NLOS scenario, we employ the APS extracted from the channel at $f_1$ to facilitate robust transmission at $f_2$. The sum-rate performance for these two terminal mobility speeds is illustrated in Fig. \ref{v60}(a) and Fig. \ref{v60}(b), respectively. The results reveal that as terminal mobility increases, the robust precoding method achieves progressively more substantial performance gains compared to the MF and RZF methods, which depend solely on instantaneous CSI. Notably, in the scenario with a user speed of 80 km/h, the sum-rate performance of both the MF and RZF methods deteriorates significantly. Furthermore, the proposed ME-RB method consistently approaches the ideal transmission performance across these mobility scenarios. These simulation results underscore the reliability and effectiveness of the proposed multi-frequency cooperative transmission method in high-mobility environments.

			\textcolor{blue}{
				\begin{figure}[!t]
				\centering
				\subfloat[Sum rate performances at ${f_2}$ with ${N_2} = 10$, the user speed ${v_{{\rm{speed}}}} = 60$ km/h. ]{\includegraphics[scale = 0.5]{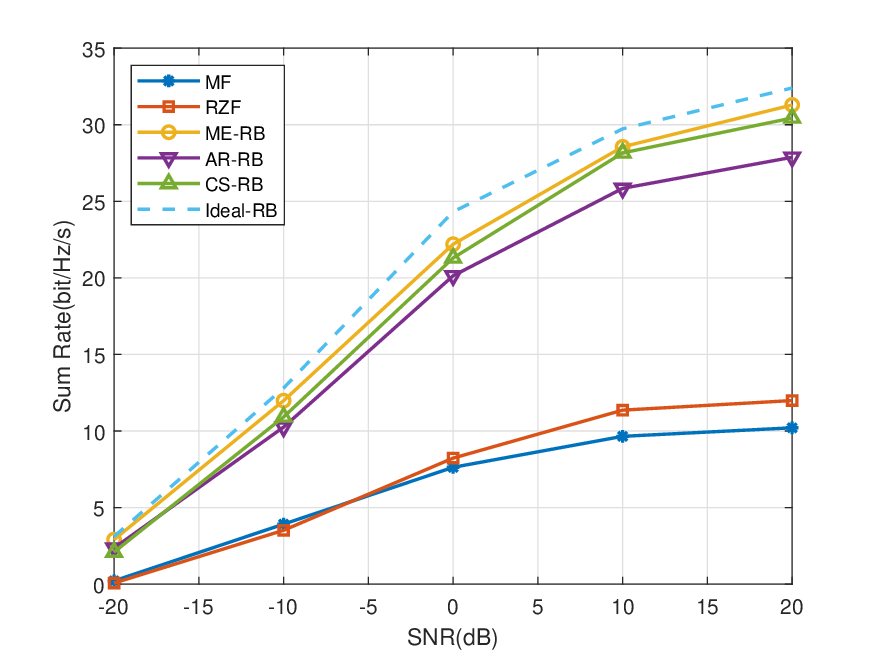}%
				}
				\\
				\subfloat[Sum rate performances at ${f_2}$ with ${N_2} = 10$, the user speed ${v_{{\rm{speed}}}} = 80$ km/h.]{\includegraphics[scale = 0.5]{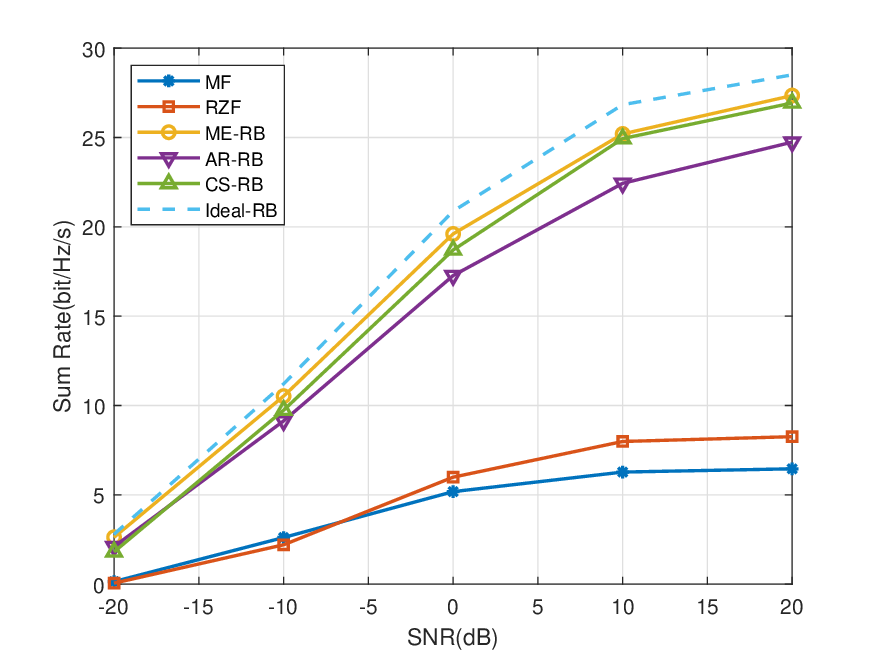}%
					\label{v80}}
				\caption{Sum rate performances of different precoding methods vesus SNR in NLOS scenario.}
					\label{v60}
			\end{figure}}

\begin{figure}[htbp]
				\centering
				\includegraphics[scale = 0.5]{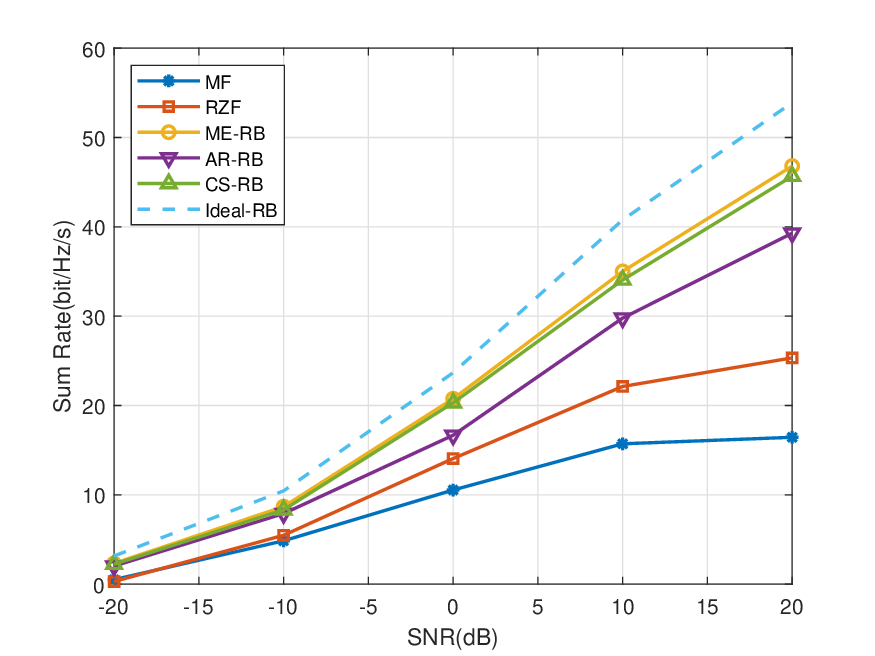}
				\caption{Sum rate performance at $f_2$ = 28GHz with $N_2$ = 16.}
				\label{mmw}
			\end{figure}

		    The proposed method is based on the assumption of high similarity in APSs across different frequency bands within the same multi-frequency system. In our earlier simulations, all frequencies were confined to the sub-6 GHz band, which aligns with this assumption and allows multi-frequency cooperative transmission to closely approach the ideal performance. However, with the increasing utilization of mmWave resources, recent studies have leveraged sub-6 GHz statistical CSI to support mmWave research by constructing multi-frequency systems. To evaluate our method in such a context, we establish a dual-frequency system with $f_1 = 3.5$ GHz, $f_2 = 28$ GHz, $N_1 = 8$, and $N_2 = 16$. The simulation, conducted under NLOS conditions with a user speed of 20 km/h, reveals that multi-frequency cooperative transmission still outperforms MF and RZF, although the gap to the ideal performance widens, as depicted in Fig. \ref{mmw}. This behavior can be attributed to the increased frequency difference leading to APS discrepancies, as discussed in \cite{9422756} and \cite{10233622}. Nevertheless, the peak power locations in the APS remain consistent across frequency bands, enabling the proposed method to retain performance advantages over MF and RZF.

			\section{Conclusion}\label{section6}

			In this paper, we addressed the acquisition of two key types of statistical CSI, i.e., spatial covariance matrices and APS, in multi-frequency massive MIMO communication systems. We first introduced a system model for multi-frequency massive MIMO and analyzed the correlation among spatial covariance matrices across different frequency bands. Building on this, we proposed an AR method to predict the covariance matrices of unknown frequency bands based on the covariance matrix of a known frequency band. Furthermore, we formulated the APS estimation problem using the ME criterion and developed a low-complexity solution by exploiting the relationship between the APS and the covariance matrix. Simulation results demonstrated the effectiveness of the proposed methods, confirming the feasibility of multi-frequency cooperative transmission, even in high-mobility and high-frequency scenarios. Notably, the proposed statistical CSI acquisition method enables channel transmission performance that closely approaches that of ideal CSI, highlighting its practical significance. 


			\appendices

			\section{Proof of Proposition \ref{prop1}}
			In order to solve  (\ref{Rb=1}), we define a set of vectors $\{ {{\bf{v}}_1},{{\bf{v}}_2}, \cdots ,{{\bf{v}}_{|{{\bar {\cal S}}_{{\rm{AR}}}}|}}\} $ in  ${\mathbb{C}}^{|{{\bar {\cal S}}_{{\rm{AR}}}}| \times 1} $, and any two of them satisfy (\ref{vRv}). Then the AR coefficient vector ${{\bf{b}}_1}$ can be regarded as a linear combination of $\{ {{\bf{v}}_1},{{\bf{v}}_2}, \cdots ,\;{{\bf{v}}_{|{{\overline {\cal S} }_{{\rm{AR}}}}|}}\} $, namely,  ${{\bf{b}}_1} =$$ \sum\limits_{i = 1}^{|{{\bar {\cal S}}_{{\rm{AR}}}}|} {{w_i}} {{\bf{v}}_i}$ for some constants ${w_1}$, $\cdots$, ${w _{{{({N_1} - 1)}^2} + 1}}$. Therefore, (\ref{Rb=1}) can be further expressed as
			\begin{equation}\label{alphaRv}
				\sum\limits_{i = 1}^{|{{\bar {\cal S}}_{{\rm{AR}}}}|} {{w _i}} {{{\bf{W}}_{\rm{\rho }}}}{{\bf{v}}_i} = \left[ {\begin{array}{*{20}{c}}
						1\\
						{\bf{0}}
				\end{array}} \right].
			\end{equation}
			Multiplying both sides of (\ref{alphaRv}) from the left with ${{\bf{v}}_1^{\rm {H}}}$, $\cdots$, ${\bf{v}}_{|{{\overline {\cal S} }_{{\rm{AR}}}}|}^{\rm{H}}$ individually, according to the property defined in (\ref{vRv}), for any $i \in \{ 1,2, \cdots ,|{\overline {\cal S} _{{\rm{AR}}}}|\}$, it can be derived that 
			\begin{equation}
				{w _i} = {\bf{v}}_i^{\rm {H}}.\left[ {\begin{array}{*{20}{c}}
						1\\
						{\bf{0}}
				\end{array}} \right] = v_i^*(0,0),
			\end{equation}
			and the linear AR coefficients of the ${N_1} \times {N_1}$ UPA model can be expressed as (\ref{b1=vv}).

			\section{Justification for the design choices of ${\alpha_d}$ and ${\beta_d}$ in (\ref{alpham}) - (\ref{betam})}
			According to the nonnegativity constraints in (\ref{tiaojian}), in the whole angle domain, the FT of $r_{k}^d(m,n)$ in ({\ref{suanfa1}}) should satisfy
			\begin{equation}\label{hengzhengtiaojian}
				\begin{array}{l}
					{{\cal{F}}}\{r_{k}^d(m,n)\}({u_i},{v_i}) = {{\cal F}\left\{ {r'(m,n)} \right\}({u_i},{v_i})} + \vspace{1ex}\\
					(1 - {\alpha _d}){\cal F}\left\{ {[{r_{k,{f_i}}}(m,n) - r'(m,n)]w(m,n)} \right\}({u_i},{v_i})  \ge 0.
				\end{array}
			\end{equation}
			Based on the property of the Fourier transform and (\ref{MEstep1}), $r'(m,n)$ is the IFT of a nonnegative function and $ {{\cal F}\left\{ {r'(m,n)} \right\}({u_i},{v_i})} \ge 0$ always holds for $\forall ({u_i},{v_i}) \in {\bf{F}}$. Therefore, when
			\begin{equation}
				\mathop {\inf }\limits_{({u_i},{v_i}) \in {\bf{F}}} \left\{ {{\cal F}\left\{ {[{r_{k,{f_i}}}(m,n) - r'(m,n)]w(m,n)} \right\}({u_i},{v_i})} \right\} < 0
			\end{equation}
			occurs for any $({{u_i}},{{v_i}})\in {\bf{F}}$, to keep (\ref{hengzhengtiaojian}),  an angular coordinate range ${\bf{F}}_{{{u_i}},{{v_i}}}^ - $ is defined as (\ref{Jxi}), and for all $({{u_i}},{{v_i}}) \in {\bf{F}}_{{{u_i}},{{v_i}}}^ -  $ we have
			\begin{equation}
				\begin{array}{l}
				(1 - {\alpha _d})\left| {{\cal F}\{ [{r_{k,{f_i}}}(m,n) - r'(m,n)]w(m,n)\} ({u_i},{v_i})} \right|\\
				\le {\cal F}\left\{ {r'(m,n)} \right\}({u_i},{v_i}),
				\end{array}
			\end{equation}
			from which it can be further derived that 
			\begin{equation}\label{alphalb}
				{\alpha _d} \ge 1 - {\alpha _{\inf }}.
			\end{equation}
		
		However, since the value of $\alpha_d$ affects whether the iterative algorithm converges, the lower bound provided in (\ref{alphalb}) may not be directly used for $\alpha_d$ \cite{1163583}. According to (\ref{suanfa1}), a smaller $\alpha_d$ indicates a smaller gap between $r_{k}^d(m,n)$ and the actual $r_{k,f_i}(m,n)$, leading to a faster convergence rate. Therefore, to avoid divergence, we ensure that ${\alpha _d} \ge {\alpha _{d - 1}}$ during the iteration. Meanwhile, we introduce the convergence rate factor ${k_d}$ in the $d$th iteration and update ${\alpha _{d}}$ as
		\begin{equation}\label{alpham1}
		{\alpha _d} = \max \left\{ {{\alpha _{d - 1}},1 - {k_d}{\alpha _{\inf }}} \right\}.
		\end{equation}
		The convergence rate of the algorithm can be adjusted by changing the value of ${k_d}$. From (\ref{alpham1}), we observe that the algorithm achieves the fastest convergence rate when ${k_d}=1$, i.e., $\alpha_d$ can be set to the minimum value satisfying the nonnegativity condition. Conversely, when ${k_d}$ approaches $0$, the convergence rate slows down accordingly.
			
			Considering the other case, when 
			\begin{equation}
			\mathop {\inf }\limits_{({u_i},{v_i}) \in {\bf{F}}} \left\{ {{\cal F}\left\{ {[{r_{k,{f_i}}}(m,n) - r'(m,n)]w(m,n)} \right\}({u_i},{v_i})} \right\} \ge 0
			\end{equation}
			is satisfied, the nonnegativity in (\ref{hengzhengtiaojian}) will always be kept, and $\alpha_d $ can take any value in the range of $[0,1]$. By setting $\alpha_d = 0$, (\ref{rkmn}) can be directly satisfied. Combining this case with (\ref{alpham1}), the final expression of ${\alpha _d}$ can be written as (\ref{alpham}).

			
		A similar method can be used to find the best value of ${\beta}_d$ in the $d$th iteration. According to (\ref{suanfa2}), since the FT of ${c^{d}_1}(m,n)$ is nonnegative, we have
		\begin{equation}\label{beta11}
			\begin{array}{l}
			{\beta _d}{\cal F}\left\{ {c_1^{d - 1}(m,n)} \right\}({u_i},{v_i}) + \vspace{1ex}\\
			(1 - {\beta _d}){\cal F}\left\{ {c'(m,n)w(m,n)} \right\}({u_i},{v_i}) \ge 0.
			\end{array}
		\end{equation}
		If ${\mathop {\inf }\limits_{({u_i},{v_i}) \in {\bf{F}}} \left\{ {{\cal F}\left\{ {c'(m,n)w(m,n)} \right\}({u_i},{v_i})} \right\} > 0}$, (\ref{beta11}) always holds for any $\beta_d\in [0,1]$, and we set $\beta_d=0$ to truncate ${c'(m,n)}$ directly.
		
		Otherwise, when $\mathop {\inf }\limits_{({u_i},{v_i}) \in {\bf{F}}} \left\{ {{\cal F}\left\{ {c'(m,n)w(m,n)} \right\}({u_i},{v_i})} \right\} $ $< 0$, (\ref{beta11}) can be further written as
		\begin{equation}
			\begin{array}{l}
			{\beta _d}\left| {{\cal F}\left\{ {c_1^{d - 1}(m,n)} \right\}({u_i},{v_i})} \right| \ge \vspace{1ex}\\
			(1 - {\beta _d})\left| {{\cal F}\left\{ {c'(m,n)w(m,n)} \right\}({u_i},{v_i})} \right|
			\end{array},
		\end{equation}
		from which the lower bound of ${\beta _d}$ can be defined as
		\begin{equation}
		\begin{array}{l}
		\beta _d^{\min } = \\
		\mathop {\sup }\limits_{({u_i},{v_i}) \in {\bf{F}}} \left\{ {\frac{{\left| {{\cal F}\left\{ {c'(m,n)w(m,n)} \right\}({u_i},{v_i})} \right|}}{{\left| {{\cal F}\left\{ {c'(m,n)w(m,n)} \right\}({u_i},{v_i})} \right| + \left| {{\cal F}\left\{ {c_1^{d - 1}(m,n)} \right\}({u_i},{v_i})} \right|}}} \right\}
		\end{array},
		\end{equation}
		i.e., ${\beta _d} \ge \beta _d^{\min }$ always holds. Similar to the analysis for ${\alpha}_d$, considering the convergence rate of the algorithm, we choose the final value of ${\beta}_d$ between ${\beta _{\min }^d}$ and 1. Regarding ${k_d}$ as the weight, ${\beta}_d$ can then be obtained as
		\begin{equation}\label{betam1}
			{\beta _{d}} = (1 - {k_d}) + {k_d}{\beta _{\min }^d}.
		\end{equation}
		Therefore, the final expression of ${\beta _d}$ can be written as (\ref{betam}).

		\bibliographystyle{IEEEtran}
		\bibliography{FINAL_VERSION}

	\end{document}